\documentclass[12pt,letterpaper]{article}

\usepackage{amsfonts}
\usepackage{amssymb}
\usepackage{amsmath}
\usepackage{amsthm}
\usepackage{multirow}
\usepackage[round,authoryear]{natbib}
\usepackage[margin=1in]{geometry}
\usepackage{booktabs}
\usepackage{dcolumn}
\newcolumntype{d}[1]{D{.}{.}{#1}} 
\usepackage{graphicx,epstopdf}
\usepackage[T1]{fontenc}
\usepackage{lmodern}
\usepackage{enumitem}
\setlist[enumerate,1]{itemsep=1pt, topsep=4pt, partopsep=0pt}
\setlist[enumerate,2]{nosep}
\setlist[itemize,1]{itemsep=1pt, topsep=4pt, partopsep=0pt}
\setlist[itemize,2]{nosep}
\usepackage{subfig}
\usepackage{bm}
\usepackage[colorlinks,linkcolor=red,citecolor=blue,pdftex,breaklinks]{hyperref}
\hypersetup{bookmarksopen=true}
\usepackage[nameinlink]{cleveref}
\usepackage{setspace}

\theoremstyle{plain}

\theoremstyle{definition}

\Crefname{assumptionx}{Assumption}{Assumptions} 
\Crefname{examplex}{Example}{Examples} 
\Crefname{remarkx}{Remark}{Remarks} 

\makeatletter
\renewcommand*{\eqref}[1]{\hyperref[{#1}]{\textup{\tagform@{\ref*{#1}}}}}
\makeatother

\makeatletter
\DeclareRobustCommand\citepos
  {\begingroup\def\NAT@nmfmt##1{{\NAT@up##1's}}%
   \NAT@swafalse\let\NAT@ctype\z@\NAT@partrue
   \@ifstar{\NAT@fulltrue\NAT@citetp}{\NAT@fullfalse\NAT@citetp}}
\makeatother

\expandafter
\def \expandafter \normalsize \expandafter{\normalsize \setlength \abovedisplayskip{10pt plus 2pt minus 7pt}}
\expandafter
\def \expandafter \normalsize \expandafter{\normalsize \setlength \abovedisplayshortskip{0pt plus 2pt}}
\expandafter
\def \expandafter \normalsize \expandafter{\normalsize \setlength \belowdisplayskip{10pt plus 2pt minus 7pt}}
\expandafter
\def \expandafter \normalsize \expandafter{\normalsize \setlength \belowdisplayshortskip{5pt plus 2pt minus 3pt}}

\def\plim{\mathop{\rm plim}}
\def\asyeq{\buildrel{a}\over{=}}
\def\bia{{\bm{a}}}
\def\bib{{\bm{b}}}
\def\bbeta{{\bm\beta}}
\def\biX{{\bm{X}}}

\def\biA{{\bm{A}}}

\def\biH{{\bm{H}}}

\def\biJ{{\bm{J}}}

\def\biX{{\bm{X}}}
\def\biR{{\bm{R}}}

\def\bir{{\bm{r}}}
\def\bis{{\bm{s}}}
\def\biy{{\bm{y}}}

\def\biu{{\bm{u}}}

\def\biw{{\bm{w}}}

\def\biV{{\bm{V}}}

\def\bgamma{{\bm{\gamma}}}

\def\bdelta{{\bm{\delta}}}

\def\bUpsilon{{\bm{\Upsilon}}}
\def\E{{\rm E}}

\def\IF{{\mathbb I}}

\def\N{{\rm N}}

\def\tk{\kern 0.08333em}
\def\tn{\kern -0.08333em}
\def\tkk{\kern 0.04167em}
\def\bzero{{\bm{0}}}
\def\bfI{{\bf I}}

\def\th#1{$#1^{\tk{\rm th}}$}

\newcommand{\FO}[1]{}

\begin{document}

\title{Cluster-Robust Jackknife and Bootstrap Inference\\for Logistic
Regression Models\thanks{MacKinnon and Webb thank the Social Sciences
and Humanities Research Council of Canada (SSHRC grant 435-2021-0396),
and Nielsen thanks the Danish National Research Foundation (DNRF Chair
grant number DNRF154), for financial support. MacKinnon and Nielsen
are also grateful for support from the Aarhus Center for Econometrics
(ACE) funded by the Danish National Research Foundation grant number
DNRF186. We thank two referees and participants at the CEA Annual
Meeting, the Canadian Econometric Study Group, the University of
Victoria, the Conference to Celebrate Professor M.~H.~Pesaran's
Achievements at U.S.C., New York Camp Econometrics, and Vanderbilt
University. An earlier version of this paper was circulated under the
title ``Cluster-robust jackknife and bootstrap inference for binary
response models.'' Code and data files may be found at
\url{http://qed.econ.queensu.ca/pub/faculty/mackinnon/logitjack/} }}

\author{James G. MacKinnon\thanks{Corresponding author. Address: 
Department of Economics, 94 University Avenue, Queen's University, 
Kingston, Ontario K7L 3N6, Canada. Email:\ \texttt{mackinno@queensu.ca}.
Tel.\ 613-533-2293. Fax 613-533-6668.}\\{\small Dept.\ of Economics, Queen's University}\\
{\small \texttt{mackinno@queensu.ca}}  \and
Morten \O rregaard Nielsen\\{\small Aarhus Center for Econometrics, Aarhus University}\\
{\small \texttt{mon@econ.au.dk}} \and
Matthew D. Webb\\{\small Dept.\ of Economics, Carleton University}\\
{\small \texttt{matt.webb@carleton.ca}}}

\maketitle

\begin{abstract}
We study cluster-robust inference for logistic regression (logit)
models. Inference based on the most commonly-used cluster-robust
variance matrix estimator (CRVE) can be very unreliable. We study
several alternatives. Conceptually the simplest of these, but also the
most computationally demanding, involves jackknifing at the cluster
level. We also propose a linearized version of the cluster-jackknife
variance matrix estimator as well as linearized versions of the wild
cluster bootstrap. The linearizations are based on empirical scores
and are computationally efficient. Our results can readily be
generalized to other binary response models. We also discuss a new
\texttt{Stata} software package called \texttt{logitjack} which
implements these procedures. Simulation results strongly favor the new
methods, and two empirical examples suggest that it can be important
to use them in practice.

\vskip 6pt

\medskip \noindent \textbf{Keywords:} logit model, logistic
regression, clustered data, grouped data, cluster-robust variance
estimator, CRVE, cluster jackknife, robust inference, wild cluster
bootstrap, linearization

\medskip \noindent \textbf{JEL Codes:} C12, C15, C21, C23.

\end{abstract}

\clearpage
\onehalfspacing

\section{Introduction}
\label{sec:intro}

Cluster-robust inference has been studied extensively over the past
decade. A recent guide to this literature is \citet*{MNW-guide}.
Other surveys include \citet{CM_2015}, \citet{JGM-CJE},
\citet{Esarey_2019}, and \citet{MW-survey}. \citet*{CGH_2018} surveys
a broader class of methods for various types of dependent data.
Although the literature has grown enormously, a very large fraction of
it concerns linear regression models estimated by ordinary least
squares. With the important exception of \citet{HansenLee_2019}, it
has largely ignored nonlinear models. For linear regression models,
several different cluster-robust variance matrix estimators (CRVEs)
are available, along with a number of bootstrap methods. The
finite\tkk-sample properties of these methods can vary greatly, and
quite a lot is known about most of them. However, there exist almost
no comparable results for nonlinear models.

To study the finite\tkk-sample properties of methods for
cluster-robust inference for nonlinear models, it is essential to
specify a particular class of such models. It seems natural to start
with binary response models because they are widely used with the sort
of cross-section and panel datasets where cluster-robust inference is
often needed. As a leading example, we focus on the logistic
regression, or logit, model.

As we show in \Cref{sec:simuls}, the only existing CRVE for logit
models that is widely used can have poor finite\tkk-sample properties.
We therefore propose several alternative procedures based on the
cluster jackknife or the wild cluster bootstrap. The first
cluster-jackknife procedures that we introduce are similar to the ones
for linear models discussed in \citet*{MNW-bootknife,MNW-influence}
and \citet{Hansen-jack}, but they are more challenging computationally
because nonlinear estimation is needed. Accordingly, we introduce
computationally simpler procedures based on score vectors at the
cluster level. These procedures, which appear to be new, involve
linearizing the first-order conditions so as to compute approximations
to the delete\tkk-one\tkk-cluster estimates needed for the
jackknife. The linearized cluster jackknife estimators appear to be
feasible for large samples with either few large clusters or many
small ones.

The same linearization methods make it possible to apply what is
essentially the wild cluster bootstrap \citep*{CGM_2008,DMN_2019} to
logistic regression models. We propose several new wild bootstrap
methods which can be computed using almost the same code as similar
wild cluster bootstrap methods for OLS regression. The methods that
seem to work best in many cases are very similar to the WCR-S and
WCU-S bootstraps proposed in \citet*{MNW-bootknife}; see
\Cref{sec:linear}.

In \Cref{sec:lrm}, we discuss sandwich CRVEs for logistic regression
models with $G$ clusters. These are special cases of the conventional
CRVEs discussed in \citet{HansenLee_2019}. We also discuss two CRVEs
based on the cluster jackknife, in which each cluster in turn is
deleted from the sample so as to obtain $G$ vectors of parameter
estimates. Although the cluster jackknife is not new, it does not seem
to have been studied in this context. Then, in \Cref{sec:linear}, we
discuss a linearization procedure and show how it can be used as the
key part of computationally efficient jackknife and wild bootstrap
procedures, which appear to be new.

In \Cref{sec:FEs}, we briefly discuss cluster fixed effects, which are
commonly encountered in models with clustered data. All the jackknife
methods need to be modified to handle them. We mainly focus on 
hypothesis tests, but \Cref{sec:CIs} discusses confidence intervals, 
where computational issues are important. In this paper we do not, 
however, discuss either predictions or marginal effects (partial 
derivatives of the logit probabilities with respect to the explanatory
variables). These topics require a more extensive treatment than we
can provide here. \Cref{sec:simuls} presents the results of a large
number of simulation experiments. \Cref{sec:examples} discusses two
empirical examples which illustrate the application of our proposed
methods. Finally, \Cref{sec:remarks} concludes.

\section{Sandwich CRVEs for Logistic Regression Models}
\label{sec:lrm}

We are concerned with the logistic regression model
\begin{equation}
\Pr\tkk(y_{gi} = 1 \,|\, \biX_{gi}) = \Lambda(\biX_{gi}\bbeta),
\quad g=1,\ldots,G, \quad i=1,\ldots,N_g.
\label{eq:lrm}
\end{equation}
Here $y_{gi}$, which equals either~0 or~1, is the response for
observation $i$ in cluster~$g$. There are $N = \sum_{g=1}^G N_g$
observations. The logistic function $\Lambda(x) = 1/(1 + e^{-x}) =
e^x/(1 + e^x)$ maps from the real line to the 0\tkk-1 interval. The
row vector $\biX_{gi}$ contains the values of $k$ explanatory
variables, and the \mbox{$k$-vector} $\bbeta$ is to be estimated. In
many cases, one element of $\bbeta$ is of particular interest, and we
wish to test a hypothesis about it or form a confidence interval.
Without loss of generality, we assume that this is the \th{k} element.
Then $\bbeta$ can be divided into a $(k-1)$-vector $\bbeta_1$ and a
scalar~$\beta_k$.

As specified in \eqref{eq:lrm}, the logistic regression model may or
may not involve any intra-cluster correlation. That will depend on
just how the $y_{gi}$ are obtained from the probabilities given by
$\Lambda(\biX_{gi}\bbeta)$; see \Cref{sec:simuls}. For the rest of
this section, we allow for the possibility that intra-cluster
correlation exists.

If $\biy$ is an $N$-vector with typical element $y_{gi}$, the
pseudo-loglikelihood function for \eqref{eq:lrm} can be written~as
\begin{equation}
\ell(\biy,\bbeta) = \sum_{g=1}^G\sum_{i=1}^{N_g}\big(
y_{gi} \log \Lambda(\biX_{gi}\bbeta) + 
(1-y_{gi})\log \Lambda(-\biX_{gi}\bbeta)\big).
\label{eq:loglik}
\end{equation}
Following \citet{HansenLee_2019}, we call \eqref{eq:loglik} a
pseudo-loglikelihood function because it assumes (incorrectly) that
the observations are independent. Using the fact that the first
derivative of $\Lambda(x)$ is $\Lambda(x)\Lambda(-x)$, the score
vector for the \th{g} cluster is simply
\begin{equation}
\bis_g(\bbeta) = \sum_{i=1}^{N_g}\bis_{gi}(\bbeta) = \sum_{i=1}^{N_g}
\big(y_{gi} - \Lambda(\biX_{gi}\bbeta)\big)\biX_{gi}.
\label{eq:scores}
\end{equation}
Thus, the first-order condition for $\hat\bbeta$ can be written as
\begin{equation}
\hat\bis =\sum_{g=1}^G \hat\bis_g = \sum_{g=1}^G \bis_g (\hat\bbeta) = 
\bzero.
\label{eq:foc}
\end{equation}
Of course, if the scores were assumed to be independent within 
clusters, it would be more natural to write $\hat\bis$ as the
summation of the $N$ empirical score vectors $\bis_{gi}(\hat\bbeta)$.
But we are merely assuming independence across clusters, with
potentially arbitrary patterns of intra-cluster dependence.

Most treatments of the logistic regression model assume that the
observations are independent or, equivalently, that each cluster
contains just one observation. In that case, the asymptotic variance
matrix is readily obtained from the result that
\begin{equation}
N^{1/2}(\hat\bbeta - \bbeta_0) \asyeq - \left(\plim N^{-1}
\biH(\bbeta_0)\right)^{\!-1} N^{-1/2} \sum_{i=1}^N \bis_i(\bbeta_0),
\label{eq:asyeq}
\end{equation}
where ``$\asyeq$'' denotes asymptotic equality, $\biH(\bbeta)$ is the
Hessian, $\bbeta_0$ is the true value of $\bbeta$, and 
$\bis_i(\bbeta_0)$ is $\bis_g(\bbeta_0)$ for the special case in which
clusters and observations coincide. For the logit model, the
information matrix is equal to minus the Hessian. Thus, in the absence
of clustering, \eqref{eq:asyeq} leads to the variance matrix estimator
\begin{equation}
\hat\biV(\hat\bbeta) = -\biH(\hat\bbeta)^{-1} 
= (\biX^\top\bUpsilon(\hat\bbeta)\biX)^{-1},
\label{eq:info}
\end{equation}
where $\bUpsilon(\bbeta)$ is an $N\times N$ diagonal matrix with 
typical diagonal element
\begin{equation}
\Upsilon_i(\bbeta) = \Lambda(\biX_i\bbeta)\Lambda(-\biX_i\bbeta);
\label{eq:ups}
\end{equation}
see, among many others, \citet[Section~11.3]{DM_2004}.

The asymptotic equality in \eqref{eq:asyeq} may not hold when there is
clustering, because the rate at which $\hat\bbeta$ tends to $\bbeta_0$
is, in general, not $N^{-1/2}$; see \citet*{DMN_2019}. When all of the 
$N_g$ are bounded, $\hat\bbeta$ does converge at the usual rate. 
However, even when some of the $N_g$ increase with~$N$, it is often
possible to make asymptotically valid inferences based on a CRVE,
because the test statistics are self-normalized. The key condition is
that all the $N_g$ must grow slowly enough so that the influence of
every cluster is asymptotically negligible.

The variance matrix of $\hat\bbeta$ may be estimated by the CRVE
\begin{equation}
\mbox{CV$_{\tn1}$:} \qquad \hat\biV_1(\hat\bbeta)
= \frac{G}{G-1}\frac{N-1}{N-k}\tk(\biX^\top  \bUpsilon(\hat\bbeta) \biX)^{-1}\!
\left(\tk\sum_{g=1}^G \hat\bis_g\hat\bis_g^\top\!\right)\!
(\biX^\top  \bUpsilon(\hat\bbeta) \biX)^{-1}.
\label{eq:CV1}
\end{equation}
This estimator is asymptotically valid as $G\to\infty$ under the 
conditions of \citet[Theorems~10 and~11]{HansenLee_2019}. It has a 
familiar sandwich form. The filling in the sandwich is the obvious 
estimator of $\E\big(\bis_g(\bbeta)\bis_g^\top(\bbeta)\big)$, and the 
bread is simply the inverse Hessian in~\eqref{eq:info}. The 
degrees-of-freedom factor is optional, but it seems reasonable to 
include it by analogy with the usual CV$_{\tn1}$ estimator for linear 
regression models. The estimator in \eqref{eq:CV1} is almost the same 
as the one used by \texttt{Stata}, which omits the factor of $(N-1)/(N-k)$.

A large body of simulation evidence suggests that the analog of
CV$_{\tn1}$ for linear regression models can perform poorly in many
cases. Recent work \citep*{MNW-bootknife,MNW-influence,Hansen-jack}
suggests that cluster-jackknife variance matrix estimators can work
much better. Each cluster is deleted in turn, yielding the vector of
delete\tkk-one\tkk-cluster estimates $\hat\bbeta^{(g)}$ when the
\th{g} cluster is deleted. The variance of the $\hat\bbeta^{(g)}$ is
then used to estimate the variance of~$\hat\bbeta$.

There are two cluster-jackknife CRVEs. One is
\begin{equation}
\mbox{CV$_{\tn3{\rm J}}$:}\qquad \hat\biV_{3{\rm J}}(\hat\bbeta) =
\frac{G-1}{G} \sum_{g=1}^G (\hat\bbeta^{(g)} -
\bar\bbeta)(\hat\bbeta^{(g)} - \bar\bbeta)^\top,
\label{eq:CV3J}
\end{equation}
where $\bar\bbeta$ is the arithmetic mean of the $\hat\bbeta^{(g)}$,
and the other is
\begin{equation}
\mbox{CV$_{\tn3}$:}\qquad \hat\biV_3(\hat\bbeta) =
\frac{G-1}{G} \sum_{g=1}^G (\hat\bbeta^{(g)} -
\hat\bbeta)(\hat\bbeta^{(g)} - \hat\bbeta)^\top,
\label{eq:CV3}
\end{equation}
which is more commonly used. It differs from \eqref{eq:CV3J} only
because it computes the variance around $\hat\bbeta$ instead 
of~$\bar\bbeta$. Strictly speaking, CV$_{\tn3}$ is estimating
the mean-squared error of $\hat\bbeta$ rather than its variance. 
Accordingly, \texttt{Stata} uses the options \texttt{vce(jackknife)} for 
CV$_{\tn3{\rm J}}$ and \texttt{vce(jackknife, mse)} for CV$_{\tn3}$.

The notation in \eqref{eq:CV3J} and \eqref{eq:CV3} is descended from
the use of HC$_3$ in \citet{MW_1985} to denote a
heteroskedasticity-consistent variance matrix estimator based on the
jackknife. \citet{BM_2002} discusses both \eqref{eq:CV3J} and
\eqref{eq:CV3} for the linear regression case but computes them in a
way analogous to HC$_3$ so that they have the usual sandwich form.
This computational method is attractive when all the $N_g$ are very
small, but it can be extremely expensive, or even infeasible, when any
of them is large \citep[Section~4]{MNW-bootknife}. Simulation evidence
in \citet{BM_2002} and \citet*{MNW-bootknife} suggests that, for
linear regression models, CV$_{\tn3{\rm J}}$ and CV$_{\tn3}$ tend to
be very similar. The former is always at least slightly smaller than
the latter, however, because the variation of the $\hat\bbeta^{(g)}$
around their mean of $\bar\bbeta$ cannot exceed their variation around
any other vector, including~$\hat\bbeta$.

The asymptotic validity of CV$_{\tn3{\rm J}}$ for a very wide variety
of models and estimators with independent observations (i.e., $N=G$)
was proved in \citet{Efron-Stein}. Similar results were proved for the
linear regression model with clustering in \citet{Hansen-jack}.
Without the factor of $(G-1)/G$, both CV$_{\tn3{\rm J}}$ and
CV$_{\tn3}$ tend to be conservative as estimators of
$\biV(\hat\bbeta)$, because they are really estimating the variance,
or mean-squared error, of statistics that are based on only $G-1$ 
clusters. However, even when cluster-jackknife standard errors are 
conservative, tests based on them are often not conservative, because 
the standard errors tend to be correlated with the parameter estimates; 
see \Cref{sec:simuls}.

It is inevitably costlier to compute CV$_{\tn3{\rm J}}$ or CV$_{\tn3}$
for a logit model than for a linear regression model with similar
numbers of parameters, clusters, and observations, because in the
former case we need to perform $G+1$ nonlinear optimizations. Much of
the time, however, $\hat\bbeta$ should provide a good starting point
for obtaining each of the $\hat\bbeta^{(g)}$. Thus the cost of
computing $G+1$ sets of estimates should be less than $G+1$ times as
great as the cost of computing $\hat\bbeta$ by itself. Moreover,
unless $G$ is extremely large, computing $G+1$ sets of estimates will
be much cheaper than any bootstrap method that requires nonlinear
estimation for every bootstrap sample. For this reason, the bootstrap
methods introduced in \Cref{sec:linear} do not require any nonlinear
estimation within the bootstrap procedure.

Another advantage of jackknife methods is that they can readily be
adapted to make inferences about smooth functions of $\bbeta$. For
example, if we care about $\delta=\beta_2/\beta_3$, we simply need to
calculate $\hat\delta$ for the entire sample and $\hat\delta^{(g)}$
for each vector of delete\tkk-one estimates and then use the analog of
\eqref{eq:CV3J} or \eqref{eq:CV3} to calculate its jackknife variance.
Bootstrap methods also have this useful feature.

The jackknife methods we propose do, however, suffer from a
potentially important computational problem. Suppose there exists some
linear combination of the $\biX_{gi}$, say $\biX_{gi}\bbeta^\bullet$,
with the property that
\begin{align}
y_{gi} &= 0\quad \mbox{whenever}\quad
\biX_{gi}\bbeta^\bullet < 0,\enspace\mbox{and}\\
y_{gi} &= 1\quad \hbox{whenever}\quad
\biX_{gi}\bbeta^\bullet > 0.
\end{align}
Then it is possible to make the value of the pseudo-loglikelihood 
function \eqref{eq:loglik}, which is always negative, arbitrarily close 
to~0 by setting $\bbeta =\gamma\bbeta^\bullet$ and letting $\gamma\to 
\infty$. This is precisely what a numerical optimization routine will 
attempt to do, although it will normally stop with an error message long
before any element of $\hat\bbeta$ becomes infinitely large. In this
case, the vector $\biX\tn\bbeta^\bullet$, which of course is not unique,
is said to be a perfect classifier, since it allows us to predict
$y_{gi}$ with 100\% accuracy for every observation in the sample.

When there is a perfect classifier, we cannot obtain well-defined
estimates of all the parameters by maximizing \eqref{eq:loglik}. If
this happens for the entire sample, then we either need to drop one or
more regressors, obtain additional data, or use some form of
regularization. The problem for the jackknife estimators is that, even
if there are no perfect classifiers for the entire sample, there might
be a perfect classifier for one or more of the subsamples. When this
happens, the values of CV$_{\tn3{\rm J}}$ and CV$_{\tn3}$ may become
extremely large and completely unreliable. Thus any program to compute
CV$_{\tn3{\rm J}}$ and CV$_{\tn3}$ needs to check whether there is a
perfect classifier when any one of the $G$ clusters is dropped. When
that happens, it should either report that the variance matrix could
not be computed or omit the offending vector(s) of delete\tkk-one
estimates and report that it has done so. In the latter case,
especially if the deleted cluster is large, CV$_{\tn3{\rm J}}$ is
likely to be more reliable than CV$_{\tn3}$, because $\bar\bbeta$ for
the reduced sample may differ noticeably from $\hat\bbeta$ for the
full sample.

It is straightforward to base inference on CV$_{\tn3}$ or
CV$_{\tn3{\rm J}}$. Suppose there are $r\ge1$ linear restrictions.
These can be written as $\biR\bbeta = \bir$, with $\biR$ an $r\times
k$ matrix and $\bir$ an $r$-vector. Tests of these restrictions are
commonly based on the Wald statistic
\begin{equation}
W(\hat\bbeta) = (\biR\hat\bbeta - \bir)^\top
(\biR\tkk\hat\biV\tn\biR^\top)^{-1} (\biR\hat\bbeta - \bir),
\label{eq:Waldstat}
\end{equation}
where $\hat\biV$ could be any of the CRVEs defined in \eqref{eq:CV1},
\eqref{eq:CV3J}, or~\eqref{eq:CV3}. Asymptotically, as $G\to\infty$,
$W(\hat\bbeta)$ is distributed as $\chi^2(r)$ under the null
hypothesis.

When there is just one restriction, the signed square root of
$W(\hat\bbeta)$ has the form of a $t$-statistic. When $\bia^\top$ is a
single row of $\biR$ and $\bir=\bzero$, such a $t$-statistic can be
written as
\begin{equation}
t_a = \frac{\bia^\top(\hat\bbeta - \bbeta_0)}
{(\bia^\top\hat\biV\bia)^{1/2}}.
\label{eq:tstat}
\end{equation}
In the very common case in which there is a single zero restriction,
say that $\beta_k=0$, \eqref{eq:tstat} reduces to $\hat\beta_k/\hat
s_k$, where $\hat s_k$ is the square root of the \th{k} diagonal
element of~$\hat\biV$\tn\tn. With linear models, it is customary to
compare $t_a$ with the $t(G-1)$ distribution \citep*{BCH_2011}.
However, both the \texttt{logit} command in \texttt{Stata} and the
\texttt{sandwich} package in \texttt{R} compare $t_a$ with the
$\N(0,1)$ distribution. This typically results in severe over-rejection,
as illustrated in Section~\ref{sec:simuls}.

\section{Methods Based on Linearization}
\label{sec:linear}

Computing either CV$_{\tn3{\rm J}}$ or CV$_{\tn3}$ requires $G+1$
nonlinear optimizations. If instead we replace the $\hat\bbeta^{(g)}$
in \eqref{eq:CV3J} and \eqref{eq:CV3} with estimates from a linear
approximation, we can obtain cluster-jackknife CRVEs that are much
cheaper to compute (\Cref{subsec:LinCJ}). We can also perform wild
bootstrap tests very inexpensively (\Cref{subsec:linboot}).

The linear approximation that we propose is based on the artificial
regression for binary response models of \citet{DM_1984}, but it does
not involve explicitly running a regression. It just uses the
contributions to the scores, $\bis_g(\bbeta)$, and to the information
matrix, $\biJ_g(\bbeta)$, made by each of the clusters. The
$\bis_g(\bbeta)$ are given by \eqref{eq:scores}, and
\begin{equation}
\biJ_g(\bbeta) = \sum_{i=1}^{N_g} \Lambda(\biX_{gi}\bbeta)
\Lambda(-\biX_{gi}\bbeta) \biX_{gi}^\top\biX_{gi},
\label{eq:infolrm}
\end{equation}
see \eqref{eq:info} and \eqref{eq:ups}. The estimates from
linearizing the model around $\bbeta$ are
\begin{equation}
\bib(\bbeta) = \bigg(\sum_{g=1}^G\biJ_g(\bbeta)\tn\bigg)^{\!\!-1}
\sum_{g=1}^G\bis_g(\bbeta) = \biJ(\bbeta)^{-1}\bis(\bbeta),
\label{eq:linearb}
\end{equation}
where $\biJ(\bbeta) = \sum_{g=1}^G \biJ_g(\bbeta)$ and $\bis(\bbeta) =
\sum_{g=1}^G \bis_g(\bbeta)$. When the $\bis_g(\bbeta)$ and
$\biJ_g(\bbeta)$ are evaluated at the true value~$\bbeta_0$, the
estimate $\bib(\bbeta_0)$ provides a linear approximation to
$\hat\bbeta-\bbeta_0$.

How well the linearization \eqref{eq:linearb} performs inevitably
depends on the model and dataset. Simulation results in
\Cref{sec:simuls} suggest that it generally performs extremely well,
except sometimes when the expectation of $y_{gi}$ is close to~0 or~1.

\subsection{The Linearized Cluster Jackknife}
\label{subsec:LinCJ}

To compute linear approximations to the delete\tkk-one\tkk-cluster
estimates, we first estimate the model by
maximizing~\eqref{eq:loglik}. Then we form the cluster-level vectors
and matrices $\hat\bis_g = \bis_g (\hat\bbeta)$ and $\hat\biJ_g =
\biJ_g (\hat\bbeta)$ using \eqref{eq:scores} and~\eqref{eq:infolrm}.
If follows from \eqref{eq:linearb} that the linear approximations to
$\hat\bbeta^{(g)} - \hat\bbeta$ when each cluster is omitted in turn
are
\begin{equation}
\hat\bib^{(g)} = (\hat\biJ - \hat\biJ_g)^{-1}
(\hat\bis - \hat\bis_g), \quad g=1,\ldots,G.
\label{eq:delone}
\end{equation}
We can use these approximations to compute cluster-jackknife variance
matrices. The one comparable to \eqref{eq:CV3} is
\vskip -6pt
\begin{equation}
\mbox{CV$_{\tn3{\rm L}}$:}\qquad \hat\biV_{3{\rm L}}(\hat\bbeta) =
\frac{G-1}{G} \sum_{g=1}^G \hat\bib^{(g)} \hat\bib^{(g)\top}.
\label{eq:CV3L}
\end{equation}
Nothing is subtracted from the $\hat\bib^{(g)}$ here, because when we
evaluate \eqref{eq:linearb} at~$\hat\bbeta$, the estimate $\hat\bib =
\bib (\hat\bbeta)$ is identically zero by the first-order conditions
for~$\hat\bbeta$. We could instead subtract $\bar\bib$, the arithmetic
mean of the $\hat\bib^{(g)}$. If we did so, we would obtain a
linearized cluster-jackknife CRVE, say CV$_{\tn3{\rm LJ}}$, comparable
to~\eqref{eq:CV3J}.

Using \eqref{eq:delone} and \eqref{eq:CV3L} to compute CV$_{\tn3{\rm
L}}$ is generally far less expensive than computing CV$_{\tn3}$. For
the empirical example of \Cref{sec:tuition}, the former is cheaper
than the latter by a factor of about forty, and they yield almost
identical results. In the simulations of \Cref{sec:simuls}, we find
that hypothesis tests and confidence intervals based on CV$_{\tn3{\rm
L}}$ are usually very similar to ones based on CV$_{\tn3}$, but not
always. When CV$_{\tn3{\rm L}}$ and CV$_{\tn3}$ differ noticeably, the
linearization \eqref{eq:linearb} is presumably not very accurate, which
suggests that no methods based on asymptotic theory may be entirely reliable.

We also find that tests and intervals based on CV$_{\tn3{\rm J}}$ and
CV$_{\tn3}$ are often indistinguishable, and similarly for
CV$_{\tn3{\rm LJ}}$ and CV$_{\tn3{\rm L}}$. Only in cases where the
number of clusters is small and cluster sizes vary greatly do the
``J'' versions of the variance matrix yield noticeably different
results from the MSE versions. However, these are cases where all four
tests over-reject and all four intervals under-cover, so it is clearly
less bad to use CV$_{\tn3}$ or CV$_{\tn3{\rm L}}$. Thus we do not
recommend the ``J'' versions of the cluster jackknife, and our
\texttt{logitjack} package does not compute them.

The linearization given by \eqref{eq:linearb} can also be used to
compute a CV$_{\tn2{\rm L}}$ variance matrix similar to the
CV$_{\tn2}$ matrix proposed in \citet{BM_2002} and referred to there
as ``bias-reduced linearization.'' These matrices are generalizations
of the HC$_2$ matrix of \citet{MW_1985}. There is more than one way to
compute them, only one of which \citep*{NAAMW_2020} is feasible for
large samples. Just how to compute CV$_{\tn2{\rm L}}$ is discussed in
\Cref{app:CV2L}. Because the simulations in \citet*{MNW-bootknife}
suggest that CV$_{\tn2}$ very rarely performs better than CV$_{\tn3}$
(although it always performs better than CV$_{\tn1}$), we do not study
CV$_{\tn2{\rm L}}$ further.

\subsection{The Linearized Wild Cluster Bootstrap}
\label{subsec:linboot}

The linear approximation \eqref{eq:linearb} can also be used to
compute new versions of the wild cluster bootstrap, which we refer to
as ``wild cluster linearized,'' or WCL, bootstraps. Like the
score bootstraps proposed in \citet{KS_2012a}, the WCL bootstraps are
based on restricted or unrestricted empirical scores. However, they
differ in one important respect from the \citet{KS_2012a} methods.
Both procedures generate bootstrap samples from empirical bootstrap
scores, but then our WCL methods multiply those bootstrap scores by
the inverse of some version of the $\biJ$ matrix, in order to mimic
the estimation step that yields empirical scores for the actual model.

We now describe the bootstrap data-generating processes. To avoid
having to give two separate results for the restricted and
unrestricted bootstraps, we let ``$\ddot x$'' denote either ``$\tilde
x$'' or ``$\hat x$'' for any $x$. In the first step, we multiply the
score vector $\ddot\bis_g$ for cluster $g$ by random variates
$v_g^{*b}$ for $b=1,\ldots,B$ bootstrap samples. The $v_g^{*b}$ must
have mean~0 and variance~1. In most cases, it seems best for them to
be independent draws from the Rademacher distribution, for which
$v_g^{*b}$ equals $+1$ and $-1$ with equal probabilities; see
\citet*{DMN_2019}. Thus the bootstrap score vectors are
\begin{equation}
\ddot\bis_g^{*b} = v_g^{*b}\ddot\bis_g, \quad g=1,\ldots,G.
\label{bootscore}
\end{equation}
The next step is to estimate the coefficient vector $\bib$ by
least squares:
\begin{equation}
\ddot\bib^{*b} = \bigg(\sum_{g=1}^G \ddot\biJ_g\!\bigg)^{\!\!-1}
\sum_{g=1}^G\ddot\bis_g^{*b}.
\label{bibddot}
\end{equation}
The vector $\ddot\bib^{*b}$ is then used to compute the empirical
bootstrap score vectors
\begin{equation}
\ddot\biw_g^{*b} = \ddot\bis_g^{*b} - \ddot\biJ_g\tk\ddot\bib^{*b}, \quad
g=1,\ldots,G.
\label{ebscores}
\end{equation}
These are what the bootstrap score vectors become after the model has
been ``estimated'' using the linearization~\eqref{eq:linearb}.

The CV$_{\tn1}$ bootstrap variance matrix can then be written as
\begin{equation}
\ddot\biV^*_b = \frac{G(N-1)}{(G-1)(N-k)}\,
\ddot\biJ^{-1}
\bigg(\sum_{g=1}^G \ddot\biw_g^{*b}(\ddot\biw_g^{*b})^\top\bigg)
\ddot\biJ^{-1},
\label{bVddotL}
\end{equation}
and the bootstrap $t$-statistic that corresponds to \eqref{eq:tstat} is
\begin{equation}
\ddot t_a^{*b} =
\frac{\bia^\top\ddot\bib^{*b}}
{(\bia^\top\ddot\biV^*_b\tk\bia)^{1/2}}.
\label{tbootL}
\end{equation}
In principle, we could instead compute a CV$_{\tn3}$ bootstrap
variance matrix, but using \eqref{bVddotL} makes the bootstrap
computations much faster. Transforming the bootstrap score vectors in
the way proposed in \citet*{MNW-bootknife} (see below) achieves much
the same effect as using CV$_{\tn3}$, but at far less computational
cost.

As usual, several different bootstrap $P$~values can be computed.
For cross-sectional models estimated by least squares, where bias is
generally not a problem, the symmetric bootstrap $P$~value is usually
appropriate. It is computed as
\begin{equation}
\hat{P}_{\rm s}^*(t_a) = \frac{1}{B} \sum_{b=1}^B 
\IF\big(|t_a^{*b}| > |t_a|\big),
\label{bootps}
\end{equation}
where $\IF(\cdot)$ denotes the indicator function. We reject the null
hypothesis for a test at level~$\alpha$ whenever $\hat{P}_{\rm
s}^*(t_a) < \alpha$. An alternative is the equal-tail bootstrap $P$
value
\begin{equation}
\hat{P}_{\rm et}^*(t_a) = \frac{2}{B} \min\left(\,\sum_{b=1}^B 
\IF(t_a^{*b} > t_a),\; \sum_{b=1}^B \big(\IF(t_a^{*b} \le t_a)\right).
\label{bootpet}
\end{equation}
Because the estimated slope coefficients for logit models tend to be
biased away from zero \citep{MS-bias}, it might be preferable to use
\eqref{bootpet} instead of \eqref{bootps} for these models. However,
we did not find any real difference between them in the experiments of
\Cref{sec:simuls}.

The WCL bootstrap methods that we have just described are the analogs
for logistic regression models of the classic wild cluster bootstrap
methods for OLS regression, which are called WCR-C and WCU-C in
\citet*{MNW-bootknife} to distinguish them from newer variants
introduced in that paper. We therefore refer to the two WCL methods as
the \mbox{WCLR-C} and \mbox{WCLU-C} bootstraps. As usual, the ``R''
and ``U'' here indicate whether the bootstrap DGP uses restricted or
unrestricted estimates. The ``-C'' denotes classic and indicates that 
the score vectors are not transformed before generating the bootstrap 
samples.

Many of the computations for WCR-C/WCU-C and \mbox{WCLR-C/WCLU-C} are
identical. For the former, everything depends on the score vector 
contributions, $\biX_g^\top\ddot\biu_g$, and the negative Hessian
matrix contributions,~$\biX_g^\top\biX_g$. For the latter, everything
depends in exactly the same way on the $\ddot\bis_g$ and the
$\ddot\biJ_g$. 

This insight shows that the WCLR and WCLU bootstraps can easily be
modified to make them analogous to the WCR-S and WCU-S bootstraps
proposed in \citet*{MNW-bootknife}. The modification involves replacing
the empirical scores $\ddot\bis$ in \eqref{bootscore} by transformed
empirical scores based on the cluster jackknife. The ``-S'' in the
names stands for ``transformed score.'' The key equations, adapted to
the present case, are
\begin{equation}
\acute\bis_g = \hat\bis_g - \hat\biJ_g\tk\hat\bib^{(g)}, \quad g=1,\ldots,G,
\label{modscoreu}
\end{equation}
for the unrestricted scores, and, assuming that the only restriction 
is~$\beta_k=0$,
\begin{equation}
\dot\bis_g = \tilde\bis_g 
- \tilde\biJ_{1g}\tk\tilde\bib_1^{(g)},
\quad g=1,\ldots,G,
\label{modscorer}
\end{equation}
for the restricted scores. Equations \eqref{modscoreu} and 
\eqref{modscorer} are, respectively, analogous to (38) and (37) in 
\citet*{MNW-bootknife}. In \eqref{modscorer}, the matrix 
$\tilde\biJ_{1g}$ contains the first $k-1$ columns of $\tilde\biJ_g$, and
the vector $\tilde\bib_1^{(g)}$ contains the first $k-1$ elements of
$\tilde\bib^{(g)}$. When there are $r < k$ linear restrictions,
\eqref{modscorer} can be replaced by a more complicated equation
analogous to (34) in \citet*{MNW-bootknife}.

Using the transformed empirical scores from \eqref{modscoreu} or
\eqref{modscorer} yields what we will call the \mbox{WCLU-S} and
\mbox{WCLR-S} bootstraps, respectively. The purpose of the
transformations is to undo the distortions of the empirical scores
caused by estimating~$\bbeta$, at least to the extent that it is
feasible to do so. This should allow the bootstrap DGP to mimic the
unknown true DGP more accurately. Simulation evidence in
\citet*{MNW-bootknife} suggests that the \mbox{WCR-S} and \mbox{WCU-S}
bootstraps can perform substantially better than the classic
\mbox{WCR-C} and \mbox{WCU-C} bootstraps in many cases. This also
seems to be true for \mbox{WCLR-S} and \mbox{WCLU-S} relative to
\mbox{WCLR-C} and \mbox{WCLU-C}; see \Cref{sec:simuls}. In particular,
confidence intervals based on \mbox{WCLU-S} perform very much better
than ones based on \mbox{WCLU-C}.

All the methods proposed in this section are implemented in the
\texttt{Stata} package \texttt{logitjack}; see \Cref{app:ljack}.

\subsection{The Linear Probability Model}
\label{subsec:LPM}

It is common to estimate a linear probability model (LPM)
instead of a logit model. In this subsection, we discuss the
relationship between the WCLR bootstraps proposed in
\Cref{subsec:linboot} and the existing WCR bootstraps applied to the
LPM. For the LPM, the first step is to run the regression
\begin{equation}
y_{gi} = \biX_{gi}\bdelta + u_{gi},
\quad g=1,\ldots,G, \quad i=1,\ldots,N_g,
\label{eq:LPM}
\end{equation}
where $u_{gi}$ is a disturbance term to be discussed below. There is
nothing to ensure that $0 \le \biX_{gi}\bdelta\le1$ in \eqref{eq:LPM}.
Nevertheless, when all the $\E(y_{gi}\,|\,\biX_{gi})$ are well away
from both~0 and~1, and all of the regressors are dummy variables,
least squares typically does yield estimated probabilities that lie in
the [0,1] interval most of the time and are quite similar to the ones
from a logit model. Thus it is often not very harmful to estimate the
LPM \eqref{eq:LPM} instead of the logistic regression
model~\eqref{eq:lrm}.

When an LPM is appropriate, the number of clusters and (for treatment
models) the number of treated clusters are both reasonably large, and
there is not too much inter-cluster variation, we might expect
inferences based on CV$_{\tn3}$, or even CV$_{\tn1}$, from
\eqref{eq:LPM} to be fairly reliable \citep*{MNW-guide}. When any of
these conditions is not satisfied, it may be safer to use some variant
of the restricted wild cluster, or WCR, bootstrap. When the Rademacher
distribution is used, the bootstrap dependent variable can take on
only two values, each with probability~$1/2$. If
$\biX_{gi}\tilde\bdelta$ denotes the \th{gi} fitted value from the
LPM, evaluated at the restricted estimates, these are
\begin{equation}
y_{gi}^* = \biX_{gi}\tilde\bdelta + (y_{gi} - \biX_{gi}\tilde\bdelta)
= y_{gi} \quad\mbox{and}\quad
y_{gi}^* = \biX_{gi}\tilde\bdelta - (y_{gi} - \biX_{gi}\tilde\bdelta)
= 2\biX_{gi}\tilde\bdelta - y_{gi}.
\label{eq:LPMdep}
\end{equation}
The first value here is just the actual value of $y_{gi}$, which is~0
or~1. But the second is either $2\biX_{gi}\tilde\bdelta$ or
$2\biX_{gi}\tilde\bdelta - 1$. Unless $\biX_{gi}\tilde\bdelta = 1/2$,
one of these numbers must always lie outside the [0,1] interval. Thus,
the $y_{gi}^*$ must look very different from the~$y_{gi}$. However,
they do have the correct expectation under the bootstrap DGP. If
$\E^* (\cdot)$ denotes expectation under the bootstrap probability
measure (that is, conditional on the sample), then
\begin{equation*}
\E^* (y_{gi}^*) = \frac12\E^*(y_{gi}) 
+ \frac12\bigl(2\biX_{gi}\tilde\bdelta - \E^*(y_{gi})\bigr)
= \biX_{gi}\tilde\bdelta.
\end{equation*}

Although the bootstrap regressand \eqref{eq:LPMdep} for the LPM may seem
rather strange, it leads to the WCR-C bootstrap score vector
\begin{equation}
\sum_{i=1}^{N_g} (y_{gi}^* - \biX_{gi}\tilde\bdelta) \biX_{gi} =
\begin{cases}
\sum_{i=1}^{N_g} (y_{gi} - \biX_{gi}\tilde\bdelta) \biX_{gi}
\text{ with prob.\ } 1/2, \\
\sum_{i=1}^{N_g} (\biX_{gi}\tilde\bdelta - y_{gi}) \biX_{gi}
\text{ with prob.\ } 1/2.
\end{cases} 
\label{eq:LPMbscore}
\end{equation}
This may be compared with the WCLR-C bootstrap score vector
from \eqref{eq:scores}:
\begin{equation}
\sum_{i=1}^{N_g} (y_{gi}^* - \tilde\Lambda_{gi}) \biX_{gi} =
\begin{cases}
\sum_{i=1}^{N_g} (y_{gi} - \tilde\Lambda_{gi}) \biX_{gi}
\text{ with prob.\ } 1/2, \\
\sum_{i=1}^{N_g} (\tilde\Lambda_{gi} - y_{gi}) \biX_{gi}
\text{ with prob.\ } 1/2. 
\end{cases}
\label{eq:lbscore}
\end{equation}
The bootstrap score vectors \eqref{eq:LPMbscore} and
\eqref{eq:lbscore} look very similar. The only difference is that the
former uses $\biX_{gi}\tilde\bdelta$ as the fitted value for
observation~$gi$, and the latter uses $\tilde\Lambda_{gi} =
\Lambda(\biX_{gi}\tilde\bbeta)$. This suggests that, when the LPM
provides a reasonably good approximation to a logit model, inferences
based on an LPM and either variant of the WCR bootstrap are likely to
be quite similar to inferences based on a logit model and the
corresponding variant of the WCLR bootstrap.

We would also expect inferences based on both variants of the WCU
bootstrap to be similar to inferences based on the corresponding
variants of the WCLU bootstrap, and inferences based on CV$_{\tn3}$
for the LPM to be similar to inferences based on both CV$_{\tn3}$ and
CV$_{\tn3{\rm L}}$ for the logit model. We will investigate these
conjectures in \Cref{sec:simuls}.

\section{Cluster Fixed Effects}
\label{sec:FEs}

It is quite common for models where cluster-robust inference is
employed to include cluster fixed effects. This creates some 
important issues, which we discuss in this section. Suppose that
$D^h_{gi}$ is a cluster dummy variable, with $D^h_{gi}=1$ whenever
$g=h$ and $D^h_{gi}=0$ otherwise. When a set of these variables is
added to the logit model \eqref{eq:lrm}, it becomes
\begin{equation}
\Pr\tkk(y_{gi}=1 \,|\, \biX_{gi}) = 
\Lambda\big(\biX_{gi}\bbeta + \sum_{h=1}^G \delta_h D^h_{gi}\big).
\label{eq:FEs}
\end{equation}
Note that $\biX_{gi}$ no longer includes a constant term, because it
would be collinear with the dummies. Thus, there are now $G+k-1$
parameters to estimate, but interest usually focuses on the vector
$\bbeta$, which now has $K = k-1$ elements.

Under standard regularity conditions, \eqref{eq:FEs} can be estimated
by maximum likelihood using the entire sample. But when cluster $h$ is
omitted, it is impossible to identify~$\delta_h$, because $D^h_{gi} =
0$ for all $g\ne h$. For linear regression models,
\citet*{MNW-bootknife} discusses how to compute cluster-jackknife
variance matrices when there are cluster fixed effects. The cheapest
and easiest method is often to partial out the fixed effects before
running either the full-sample regression or any of the
delete\tkk-one\tkk-cluster regressions. But this partialing-out method
is not feasible for \eqref{eq:FEs} because it is nonlinear in the
fixed effects.

A feasible method, also discussed in \citet*{MNW-bootknife}, is to use
a generalized inverse. For a linear regression model, this sets the
coefficient $\delta_h$ to~0 for the regression that omits cluster~$h$,
and $\hat\bbeta^{(h)}$ is the same as it would be for the
partialing-out method. This method can also be used for CV$_{\tn3}$ or
CV$_{\tn3{\rm J}}$, provided the logit estimation routine employs a
generalized inverse and sets the estimates of unidentified
coefficients to zero, as the ones in \texttt{R} and \texttt{Stata} do.
The generalized-inverse method is particularly easy to use for the
linearization methods proposed in \Cref{sec:linear}. We simply replace
the ordinary inverse in \eqref{eq:delone} and \eqref{bibddot} by an
appropriate generalized inverse.

A third method would be to estimate $G+1$ different logit models. The
model for the full sample would have $K+G$ coefficients, but the model
for each of the delete\tkk-one\tkk-cluster samples would have only
$K+G-1$ coefficients, because the fixed-effect dummy for the deleted
cluster must be omitted. This is conceptually straightforward, but it
may be challenging to program efficiently, because the set of fixed
effects will be different for each of the $G+1$ models.

For both feasible methods, cluster-jackknife variance matrices can be
computed in the usual way only for the vector $\hat\bbeta$, which has
$K=k-1$ coefficients. This is sufficient for inference about slope
coefficients. However, it is insufficient for inference about
predicted probabilities or marginal effects, because the constant term
for every observation in cluster~$g$ is~$\delta_g$. We need the full
variance matrix for all $G+K$ coefficients to obtain the standard
error of $\biX_{gi}\hat\bbeta + \hat\delta_g$ for any 
observation~$gi$, from which we can then compute the standard error of
the predicted probability, $\Lambda(\biX_{gi}\hat\bbeta
+\hat\delta_g)$, using the delta method.

We also need the full variance matrix in order to obtain the standard
errors of the marginal effects. Unfortunately, since each of the
$\hat\delta_g$ is identified by the observations in just one cluster,
there is no obvious way to estimate that matrix reliably. We have just
seen that none of the full CV$_{\tn3}$ variance matrices can be
computed. If instead we try to use CV$_{\tn1}$, the elements
corresponding to the $\delta_g$ will be severely biased downwards,
because each of the fixed-effect dummy variables is simply a treatment
dummy for a single treated cluster; see \citet{MW-JAE,MW-EJ}. In the
remainder of this paper, we assume for simplicity that there are no
cluster fixed effects.

\section{Confidence Intervals}
\label{sec:CIs}

There are many ways to construct confidence intervals for logistic
regression models. Some of these are computationally convenient, but
others are inconvenient because the model is nonlinear. In this
section, we briefly discuss a number of methods. The performance of
several intervals will be studied in \Cref{sec:simuls}.

The simplest approach to constructing a $100(1-\alpha)\%$ confidence
interval, where $\alpha$ often equals either 0.05 or 0.01, is to
employ a symmetric interval of the form
\begin{equation}
\big[\hat\beta_j - c_{1-\alpha/2}\tk{\rm se}(\hat\beta_j), \;\;
\hat\beta_j + c_{1-\alpha/2}\tk{\rm se}(\hat\beta_j)\big],
\label{CIpm}
\end{equation}
where $\hat\beta_j$ is the maximum likelihood estimate of the 
coefficient of interest, and $c_{1-\alpha/2}$ is the
\mbox{$1-\alpha/2$} quantile of some distribution. The critical value
$c_{1-\alpha/2}$ might come from either the $\N(0,1)$ distribution or
the $t(G-1)$ distribution, and the standard error might come from any
of several different cluster-robust variance estimators or numerous
different bootstrap distributions. It seems odd to use quantiles of
the $\N(0,1)$ distribution, which is the default for logit models in
\texttt{Stata} and \texttt{R}, when quantiles of the $t(G-1)$
distribution are usually employed to construct intervals like
\eqref{CIpm} for linear regression models using cluster-robust
standard errors. The results in \Cref{sec:simuls} suggest that the
latter is always a better choice for logit models too.

Instead of using CV$_{\tn1}$ or CV$_{\tn3}$ standard errors, we can
use a bootstrap standard error based on $B$ bootstrap 
estimates,~$\hat\beta_j^{*b}$. This is simply
\begin{equation}
{\rm se_{\tk boot}}(\hat\beta_j) = \left(\frac1{B-1} 
\sum_{b=1}^B \big(\hat\beta_j^{*b} -
\bar\beta_j^*\big)^{\!2}\right)^{\!\!1/2}\!,
\label{bootse}
\end{equation}
where $\bar\beta_j^*$ is the arithmetic mean of 
the~$\hat\beta_j^{*b}$. Any bootstrap DGP that does not impose the
null hypothesis can be used to generate the bootstrap samples.
However, using the best-known such DGP, namely, the pairs cluster
bootstrap, would be extremely expensive, because it would involve
estimating a nonlinear model for each of $B$ bootstrap samples. In
contrast, the wild cluster linearized bootstrap methods proposed in
\Cref{sec:linear} are inexpensive when the computational tricks of
\citet*{RMNW} are employed. In principle, either WCLU-C or WCLU-S could
be used, but the latter seems to work much better; see
\Cref{sec:simuls}.

Instead of using a WCLU bootstrap to estimate a bootstrap standard
error from \eqref{bootse}, we could construct a studentized bootstrap
interval of the form
\begin{equation}
\big[\hat\beta_j - c^*_{1-\alpha/2}\tk{\rm se}_1(\hat\beta_j), \;\;
\hat\beta_j - c^*_{\alpha/2}\tk{\rm se}_1(\hat\beta_j)\big].
\label{studboot}
\end{equation}
Here ${\rm se}_1(\hat\beta_j)$ is the CV$_{\tn1}$ standard error of
$\hat\beta_j$, and $c^*_{\alpha/2}$ and $c^*_{1-\alpha/2}$ are the
$\alpha/2$ and $1-\alpha/2$ quantiles of the distribution of the
bootstrap $t$-statistics. For example, if $B=999$ and $\alpha=0.05$,
these would be numbers 25 and 975 in the list of bootstrap
$t$-statistics sorted from smallest to largest. It may seem odd to use
the CV$_{\tn1}$ standard error in \eqref{studboot}, because we have
argued in \citet*{MNW-bootknife} that the CV$_{\tn3}$ standard error is
more reliable. But it is essential to use the same standard error in
\eqref{studboot} as in the WCLU bootstrap itself. The advantages of
using cluster-jackknife standard errors apply to the WCLU-S bootstrap
through the transformation \eqref{modscoreu} of the bootstrap scores.
This suggests that intervals based on WCLU-S should outperform ones
based on WCLU-C.

In theory, the studentized bootstrap interval \eqref{studboot} may
perform better than the interval \eqref{CIpm} using bootstrap standard
errors, for the same bootstrap DGP, because the former is based on a
test statistic that is asymptotically pivotal and allows the
$t$-statistic to have an asymmetric distribution. In contrast, the
latter is not based on an asymptotically pivotal quantity and imposes
symmetry on the distribution. We shall investigate this conjecture, and 
others, in \Cref{sec:simuls}.

Yet another way to obtain a bootstrap confidence interval is to invert
a bootstrap test based on a restricted bootstrap DGP, such as the
WCLR-S bootstrap. This is quite easy for linear regression models, but
a different set of bootstrap samples is needed every time we calculate
a bootstrap $P$~value. This means that, to obtain a WCLR-S confidence
interval, the logit model has to be estimated many times subject to
the restriction that $\beta_j$ equals each candidate value for the
limits of the interval; see \citet[Section~3.4]{JGM-fast}. When we
attempted to implement this method, we occasionally encountered
numerical problems in the logit routine. Although the procedure worked
most of the time, it was infeasible to perform simulations with a
large number of replications. We therefore decided not to include this
sort of interval in our simulations, and we cannot recommend it.

Based on the simulation results in \Cref{sec:simuls}, there are four
confidence intervals that we can recommend. The simplest is the
conventional interval \eqref{CIpm} based on quantiles of the $t(G-1)$
distribution and CV$_{\tn3{\rm L}}$ standard errors. Using CV$_{\tn3}$
standard errors instead works a bit better in some cases, but it can
be much more expensive. Two intervals based on the WCLU-S bootstrap
generally work well and are not expensive to compute. One is the
studentized bootstrap interval \eqref{studboot}, and the other is the
interval \eqref{CIpm} based on quantiles of the $t(G-1)$ distribution
and bootstrap standard errors from~\eqref{bootse}. Methods based on
the WCLU-C bootstrap often work much less well and are not
recommended.

\section{Simulation Evidence}
\label{sec:simuls}

We have performed a large number of simulation experiments for most of
the tests and confidence intervals discussed above. How well they
perform inevitably depends on many features of the model and DGP. In
the following subsections, we consider several specific, relevant 
scenarios in which we investigate the performance of the various 
methods.

Several interesting regularities emerge from our experiments. In 
particular, the classic CV$_{\tn1}$-based $t$-test using critical
values from the $\N(0,1)$ distribution greatly over-rejects compared
to one using the same test statistic and the $t(G-1)$ distribution.
Moreover, even the latter $t$-test is prone to over-reject, often
severely, and it almost always does so to a greater extent than the
jackknife and bootstrap tests proposed in \Cref{sec:lrm,sec:linear}.
Likewise, confidence intervals based on CV$_{\tn1}$ standard errors
are prone to under-cover much more severely than the jackknife and
WCLU-S bootstrap ones discussed in \Cref{sec:CIs}. It seems to be rare
for the better methods to yield inferences that differ substantially
from each other, but this can sometimes happen. In
\Cref{sec:examples}, we provide some advice about how to proceed when
alternative tests yield differing inferences.

\subsection{Simulation design}

In order to investigate the finite\tkk-sample properties of
cluster-robust $t$-tests and confidence intervals, we need to generate
samples with intra-cluster correlation. In principle, this could be
done in many different ways. The one that we use is particularly easy
to implement, since it just requires a uniform random number
generator. First, we specify a parameter $\phi$ between 0 and~1, which 
determines the extent of within-cluster correlation.  Then
we generate $G$ independent random variates $v_g\sim {\rm U}(0,1)$,
$N$ independent random variates $e_{gi}\sim {\rm U}(0,1)$, and up to
$N$ more independent random variates $v_{gi}\sim {\rm U}(0,1)$. For
all $g=1,\ldots,G$ and $i=1,\ldots,N_g$, we then compute 
\begin{align}
\label{eq:ugi}
u_{gi} &= v_g \mbox{ if } e_{gi} \le \phi, \mbox{ and }
u_{gi} = v_{gi} \mbox{ if } e_{gi} > \phi.\\
\label{eq:ygi}
y_{gi} &= 0 \mbox{ if } \Lambda(\biX_{gi}) \le u_{gi}, \mbox{ and }
y_{gi} = 1 \mbox{ if } \Lambda(\biX_{gi}) > u_{gi}.
\end{align}
Thus, with probability $\phi$, the random variate $u_{gi}$ is equal to
$v_g$, and, with probability \mbox{$1-\phi$}, it is equal to~$v_{gi}$.
At one extreme, when $\phi=0$, all of the $u_{gi}$ are independent. At
the other extreme, when $\phi=1$, they all take the same value~$u_g$.
The value of the binary variate $y_{gi}$ is then equal to~0 with
probability $1 - \Lambda(\biX_{gi}\bbeta)$ and to~1 with probability
$\Lambda(\biX_{gi}\bbeta)$, as usual, but these events are not 
independent across observations within each cluster unless~$\phi=0$.

Most of our experiments deal with tests of a restriction on one
parameter, which can be thought of as the coefficient on a treatment
dummy. The function $\Lambda(\biX_{gi}\bbeta)$ is given by
\begin{equation}
\Lambda\Big(\!\beta_1 + \sum_{j=2}^{k-1} \beta_j X_{gij} + \beta_k
T_{gi}\!\Big),
\label{eq:logitmodel}
\end{equation}
where the $X_{gij}$ are binary random variables. For each $j$ and for 
each~$g$, a probability $\omega_g$ between 0.25 and 0.75 is chosen at 
random for each replication. Then, with probability $\omega_g$, we set 
$X_{gij}=1$ for all $i=1,\dots ,N_g$, and otherwise we set $X_{gij}=0$.
This design is intended to mimic the situation, often encountered in 
treatment regressions, where all of the regressors are dummies. It 
allows these variables to vary moderately across clusters. In most 
experiments, $\beta_j=1$ for $1<j<k$. The model would fit better (worse)
if these coefficients were larger (smaller). The treatment regressor 
$T_{gi}$ equals 1 for $G_1$ randomly chosen clusters and~0 for the 
remaining $G_0=G-G_1$ clusters, with $\beta_k=0$ in most experiments. 
The unconditional expectation of $y_{gi}$ is~$\pi$, which depends on 
the $\beta_j$ and the distribution of the~$X_{gij}$. When we vary it, 
we do so by changing~$\beta_1$, the constant term. 

The $N$ observations are divided among the $G$ clusters using the
formula
\begin{equation}
N_g = \left\lfloor N \frac{\exp(\gamma g/G)}{\sum_{j=1}^G \exp(\gamma
j/G)}\right\rfloor\!, \quad g=1,\ldots, G-1,
\label{eq:gameq}
\end{equation}
where $\lfloor x\rfloor$ means the integer part of~$x$. The value of
$N_G$ is then set to $N - \sum_{g=1}^{G-1} N_g$. This procedure has
been used in \citet{MW-JAE}, \citet*{DMN_2019}, and several other
papers. The key parameter here is $\gamma$, which determines how
uneven the cluster sizes are. When $\gamma=0$ and $N/G$ is an integer,
\eqref{eq:gameq} implies that $N_g = N/G$ for all~$g$. For $\gamma>0$,
cluster sizes vary more and more as $\gamma$ increases. The largest
value that we use is~4. In that case, when $G=24$ and $N=12000$, the
largest cluster (1889 observations) is about 47 times as large as the
smallest (40 observations). In many of our experiments, $\gamma=2$,
which implies that the largest cluster (1120 observations) is just
under seven times as large as the smallest (163 observations).

\subsection{Canonical case}

\begin{figure}[tb]
\begin{center}
\caption{Rejection frequencies for tests at the 0.05 level
as functions of $G$}
\label{fig:A}
\includegraphics[width=0.96\textwidth]{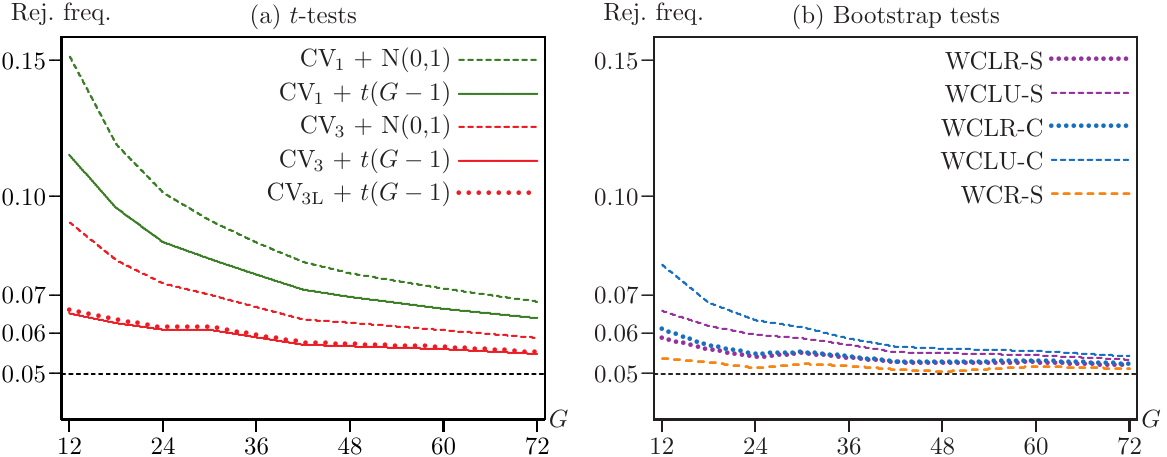}
\end{center}
{\footnotesize
\textbf{Notes:} These experiments use $100,\tn000$ replications, with
$G=12, 18, 24, 30, 36, 42, 48, 60, 72$, and $N=500G$. The value of $G_1$
is~$G/3$. There are 6 regressors, one of which is a treatment dummy
that is assigned at random, plus a constant term. The value of $\phi$
is~$0.1$. The extent to which cluster sizes vary is determined by the
parameter $\gamma$ in \eqref{eq:gameq}, which equals~2. The
unconditional expectation of $y_{gi}$ is $\pi=0.31$. CV$_{\tn1}$ and
CV$_{\tn3}$ denote cluster-robust $t$-statistics based on
\eqref{eq:CV1} and \eqref{eq:CV3}, respectively. Bootstrap tests use
$B=999$.}
\end{figure}

In the first set of experiments, we let $N$ vary from $6,\tn000$ to
$36,\tn000$, with $G=N/500$, $\gamma=2$, $\phi=0.1$, and $G_1 = G/3$.
This is not an ideal case, because the cluster sizes vary considerably, 
there is some intra-cluster correlation, and the fraction of treated 
clusters differs noticeably from one-half. However, it is a canonical 
case that seems representative of many empirical applications, and it 
is a case where we would expect most good methods to work quite well,
at least for the larger values of~$G$.

Panel~(a) of \Cref{fig:A} shows rejection frequencies as functions of
$G$ for five $t$-tests. The vertical axis has been subjected to a
square root transformation in order to handle the wide range of
observed rejection frequencies. The results in this figure are
striking. The most reliable $t$-tests use CV$_{\tn3}$ or 
CV$_{\tn3{\rm L}}$ standard errors and $t(G-1)$ critical values. They 
both reject about 6.5\% of the time when $G=12$ and 5.5\% when $G=72$. 
In contrast, the test based on CV$_{\tn1}$ standard errors and $t(G-1)$ 
critical values rejects between 6.4\% and 11.4\% of the time.

Panel~(a) also shows results for tests based on $\N(0,1)$ critical
values. We report these because, as of Version~19, \texttt{Stata}
reports $P$~values and confidence intervals based on the $\N(0,1)$
distribution for logit models, even though it reports ones based on
the $t(G-1)$ distribution for linear regression models. The
\texttt{sandwich} package in \texttt{R} does the same thing. Using
standard normal critical values necessarily yields higher rejection
frequencies than using $t(G-1)$ critical values, and the additional
over-rejection caused by using the former is not at all negligible,
especially for smaller values of~$G$. In the remaining experiments, we
only use $t(G-1)$ critical values.

Panel~(b) of \Cref{fig:A} shows rejection frequencies for five
bootstrap tests. These are all based on the symmetric $P$~value
\eqref{bootps}; results for the equal-tail $P$~value \eqref{bootpet}
were almost identical. The three restricted bootstrap tests all work
better than any of the $t$-tests. For small values of~$G$, the WCLU-S
test over-rejects about as much as the two cluster-jackknife $t$-tests
with $t(G-1)$ critical values, and the WCLU-C test over-rejects
noticeably more. For the largest values of~$G$, however, both these
tests reject less frequently than the best $t$-tests.

\begin{figure}[tb]
\begin{center}
\caption{Coverage for 95\% confidence intervals as functions of $G$}
\label{fig:B}
\includegraphics[width=0.96\textwidth]{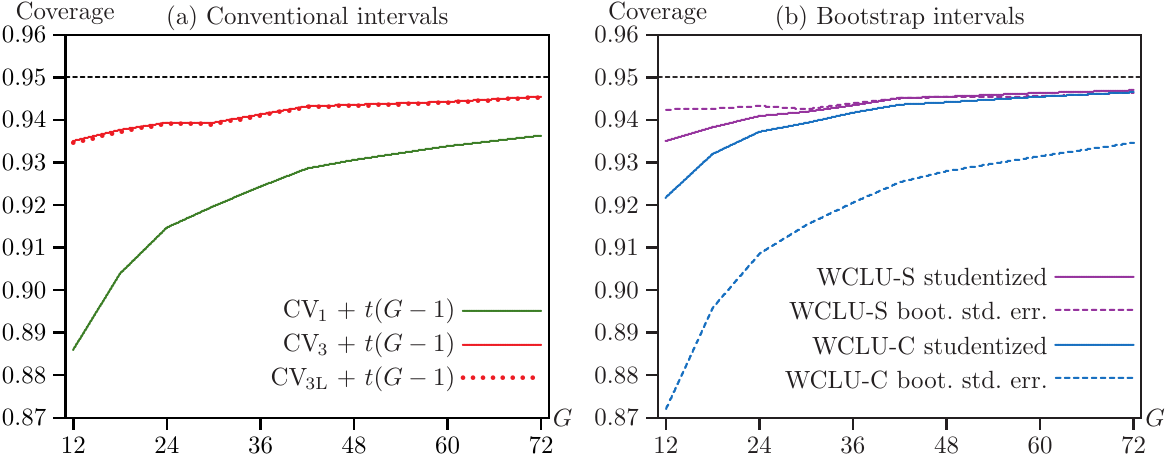}
\end{center}
{\footnotesize
\textbf{Notes:} These experiments are identical to the ones in
\Cref{fig:A}. See the notes to that figure.}
\end{figure}

\Cref{fig:B} reports coverage for confidence intervals as functions
of~$G$ based on the same experiments as in \Cref{fig:A}. Most of these
results could have been anticipated. Tests that over-reject moderately
lead to intervals that under-cover moderately. The intervals based on 
CV$_{\tn3}$ and CV$_{\tn3{\rm L}}$ standard errors perform well, as do
the ones based on the WCLU-S bootstrap. Interestingly, the studentized
bootstrap WCLU-C interval based on \eqref{studboot} covers nearly as
well as the two WCLU-S intervals, at least for larger values of~$G$,
but the interval that uses WCLU-C standard errors based on
\eqref{bootse} under-covers quite badly even when $G$ is large. For
larger values of~$G$, the two WCLU-S bootstrap intervals work only
slightly better than the two intervals based on cluster-jackknife
standard errors. For $G\ge42$, they all cover the true value more than
94\% of the time.

In the experiments reported so far, we have chosen the parameters
of the DGP to make inference at least moderately difficult. As a 
benchmark case, we next consider a set of experiments in which we 
deliberately make inference as easy as possible without making the 
number of clusters so large that asymptotic approximations are bound 
to be very good. There are 50 clusters, between 20 and 30 of which are 
treated, with all other parameters chosen to make inference as easy as 
possible. Because many methods work well, we use $400,\tn000$ 
replications in order to make it easier to distinguish among them.

\begin{figure}[tb]
\begin{center}
\caption{Rejection frequencies for tests at the .05 level in almost ideal case}
\label{fig:I}
\includegraphics[width=0.96\textwidth]{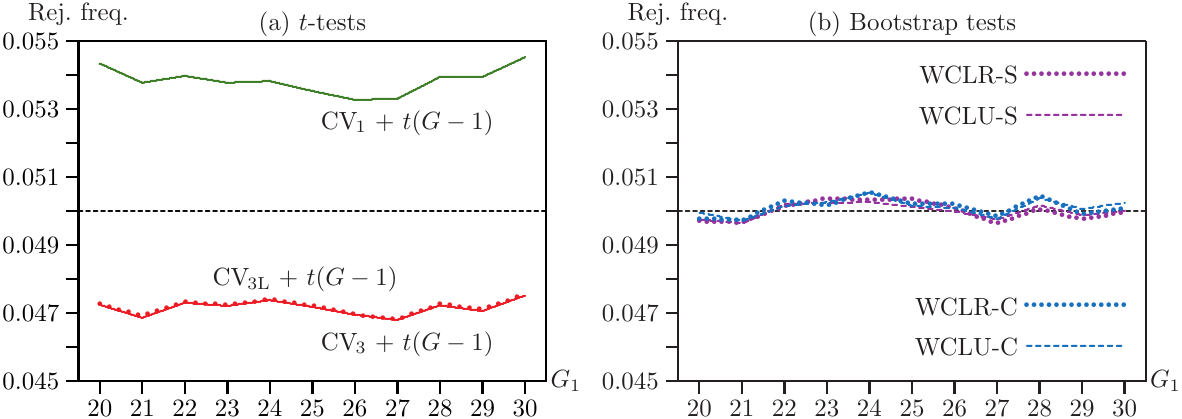}
\end{center}
{\footnotesize
\textbf{Notes:} In these experiments, $G=50$, $N=25,\tn000$,
$\gamma=0$ (so that all clusters are the same size), $\pi=0.5$, $k=7$,
and $\phi=0$ (so that there is no intra-cluster correlation). There
are $400,\tn000$ replications.}
\end{figure}

\Cref{fig:I} shows rejection frequencies for this case as functions 
of~$G_1$. In Panel~(a), we see that $t$-tests based on CV$_{\tn1}$
standard errors and the $t(49)$ distribution over-reject very
slightly, while ones based on cluster-jackknife standard errors
under-reject very slightly. For linear regression models, it is not
uncommon for the latter to under-reject a little bit in very regular
cases; see \citet*[Figure~5]{MNW-bootknife}. In Panel~(b), we see that
all the bootstrap methods perform essentially perfectly. The minor
observed deviations between their rejection rates and 0.05 could well
be due to experimental randomness. The figure does not show results
for WCR-S, because they are almost indistinguishable from the ones
for the other restricted bootstrap methods.

\subsection{Variation in treatment allocation}

\begin{figure}[tb]
\begin{center}
\caption{Rejection frequencies for tests at the 0.05 level
as functions of $G_1$}
\label{fig:C}
\includegraphics[width=0.96\textwidth]{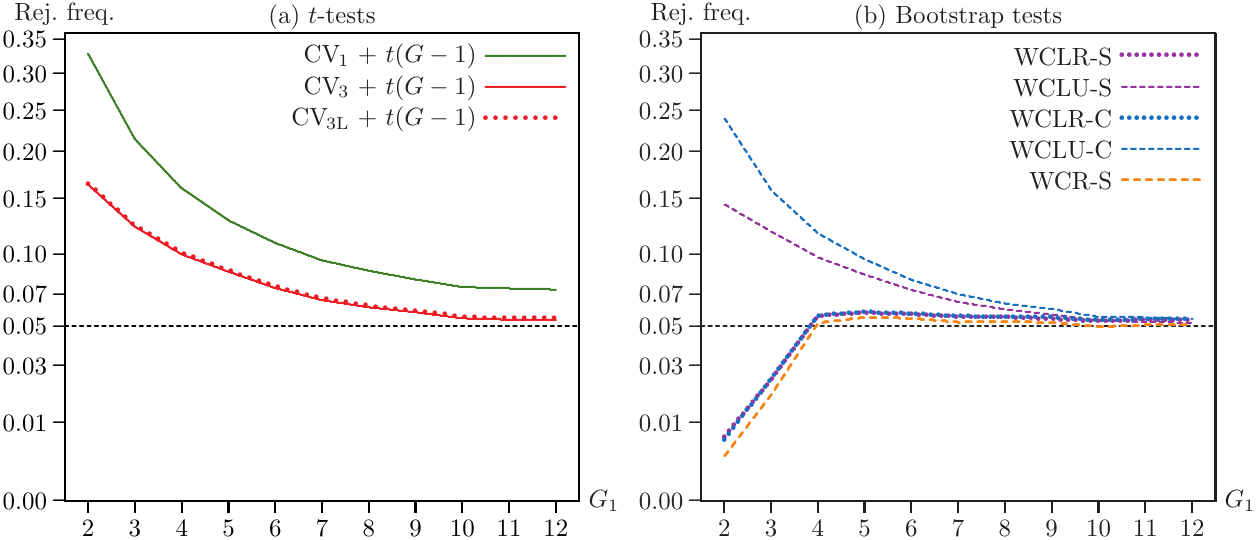}
\end{center}
{\footnotesize
\textbf{Notes:} These experiments are similar to the ones in
\Cref{fig:A}, except that $G=24$, and $G_1$ varies between 2 and~12. As
before, $\pi=0.31$, $\phi=0.1$, $\gamma=2$, and $k=7$.
There are $100,\tn000$ replications.}
\end{figure}

The next set of experiments, reported in \Cref{fig:C}, focuses on
$G_1$, the number of treated clusters. In all cases, $G=24$ and
$N=12,\tn000$, but $G_1$ varies between 2 and 12. The smallest value 
is~2, because methods based on the cluster jackknife (including the
WCLR/WCLU-S bootstraps) cannot handle the case where $G_1=1$, since
the coefficient $\beta_k$ is not identified when the single treated
cluster is omitted. The largest value is $G_1=G/2=12$, because, with
clusters treated at random, results must be symmetric in $G_1$ around 
the value~$G/2$.

In both panels of \Cref{fig:C}, we see that the performance of all
tests improves as $G_1$ increases to~$G/2$. In Panel~(a), we see that
$t$-tests based on CV$_{\tn3}$ and CV$_{\tn3{\rm L}}$ perform 
identically, and much better than ones based on CV$_{\tn1}$, although
all the $t$-tests over-reject severely for smaller values of~$G_1$. In
Panel~(b), the unrestricted bootstrap tests over-reject severely for 
smaller values of~$G_1$. The best of them, WLCU-S, performs only a 
little better than the cluster-jackknife $t$-tests. In contrast, all
the restricted bootstrap tests under-reject severely for $G_1=2$ and
$G_1=3$ but perform very well for $G_1\geq 4$. WCR bootstrap tests for
linear regression models are well known to behave in exactly the same
way; see \citet{MW-JAE,MW-EJ} for an explanation. As in \Cref{fig:A},
the best test for most values of $G_1$, by a small margin, seems to be
WCR-S. In practice, it would probably be wise to compare
$P$~values from several bootstrap methods.

\begin{figure}[tb]
\begin{center}
\caption{Rejection frequencies for tests at the 0.05 level
as functions of $\pi$}
\label{fig:E}
\includegraphics[width=0.96\textwidth]{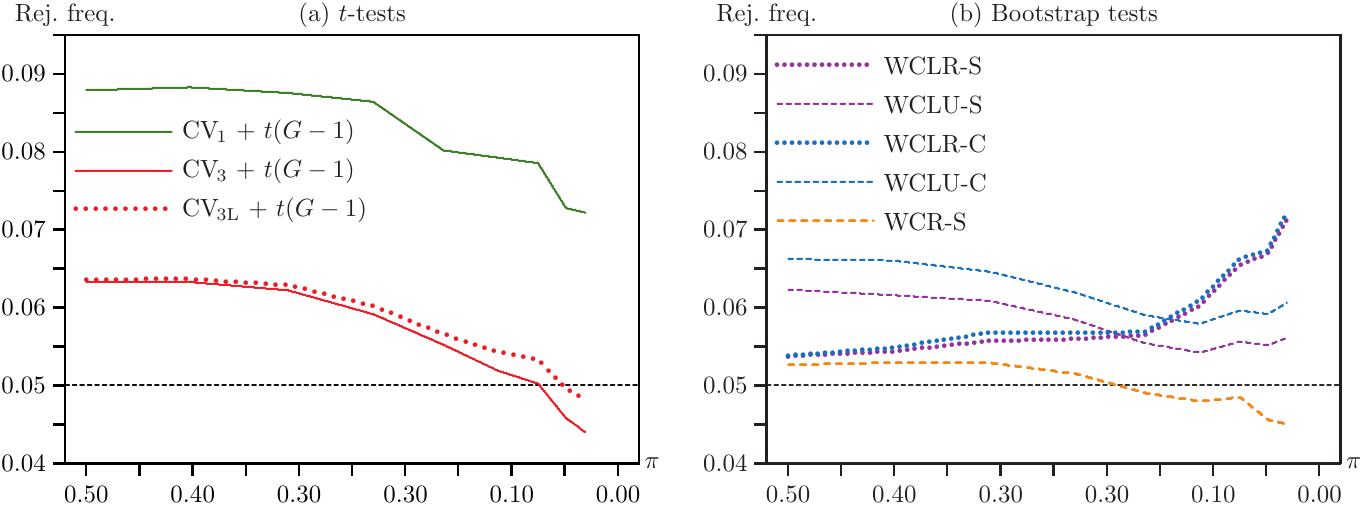}
\end{center}
{\footnotesize
\textbf{Notes:} These experiments are similar to the ones in
\Cref{fig:C}, except that $G_1=8$ and $\pi$ (the unconditional 
expectation of $y_{gi}$) varies between 0.03 and~0.50. As before, 
$G=24$, $N=12,\tn000$, $\phi=0.1$, $\gamma=2$, and $k=7$. There are
$100,\tn000$ replications.}
\end{figure}

The finite\tkk-sample properties of estimators and test statistics in
binary response models often depend on how close the average value of
the dependent variable is to one\tkk-half. Therefore, in the next set
of experiments, we vary~$\pi$, the unconditional expectation
of~$y_{gi}$, by changing the value of $\beta_1$
in~\eqref{eq:logitmodel}. In \Cref{fig:E}, the horizontal axis shows
the value of~$\pi$, which decreases from 0.50 to about 0.03 as we move
from left to right on the horizontal axis. The results must be symmetric
in $\pi$ around the value~$0.5$.

In Panel~(a), all the $t$-tests over-reject less frequently as $\pi$
decreases, with the two cluster jackknife tests eventually
under-rejecting slightly. The differences between the latter tests
now become noticeable for small values of~$\pi$. In Panel~(b),
several odd things happen. The two WCLU bootstrap tests over-reject
less often as $\pi$ decreases, at least up to a point, while the two
WCLR bootstrap tests over-reject more often. The WCR-S test actually
under-rejects for small values of~$\pi$. This figure suggests that
there may be important discrepancies between the various tests when
most  of the values of the dependent variable are either~0 or~1.

\subsection{Variation in cluster sizes and correlation}

\begin{figure}[tb]
\begin{center}
\caption{Rejection frequencies for tests at the .05 level as functions
of~$\gamma$}
\label{fig:D}
\includegraphics[width=0.96\textwidth]{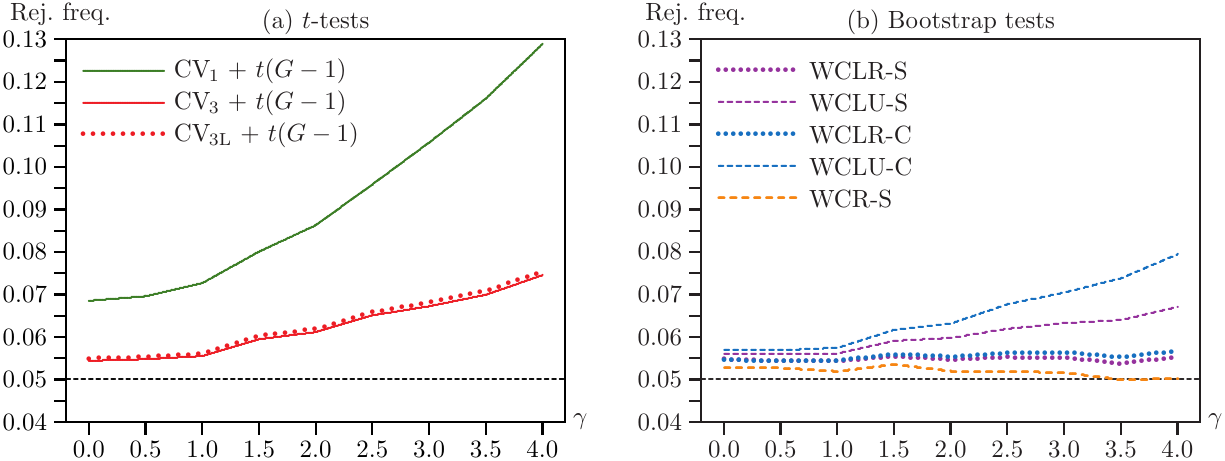}
\end{center}
{\footnotesize
\textbf{Notes:} These experiments are similar to the ones in
\Cref{fig:C}, except that $G_1=8$, and $\gamma$ varies between
0.0 and~4.0. When $\gamma=0$, all clusters have 500 observations. When
$\gamma=4$, cluster sizes range from 40 to~1889.}
\end{figure}

\Cref{fig:D} deals with the effects of cluster size variability, with
$\gamma$ varying between~0 (all cluster sizes equal~500) and~4
(cluster sizes vary greatly) on the horizontal axis. In Panel~(a), the
CV$_{\tn1}$ $t$-test always rejects substantially more often than any
of the other tests, and it does so to a greater extent as $\gamma$
increases. The CV$_{\tn3}$ and CV$_{\tn3{\rm L}}$ $t$-tests perform 
much better, but they also over-reject somewhat more frequently 
as~$\gamma$ increases.

In Panel~(b) of \Cref{fig:D}, the two WCLU bootstrap tests over-reject
more often as $\gamma$ increases, while the WCR-S test and the two
WCLR bootstrap tests reject at about the same rate for all values of
$\gamma$ considered. The latter three tests all perform very well
throughout the figure.

\begin{figure}[tb]
\begin{center}
\caption{Rejection frequencies for tests at the 0.05 level
as functions of $\phi$}
\label{fig:F}
\includegraphics[width=0.96\textwidth]{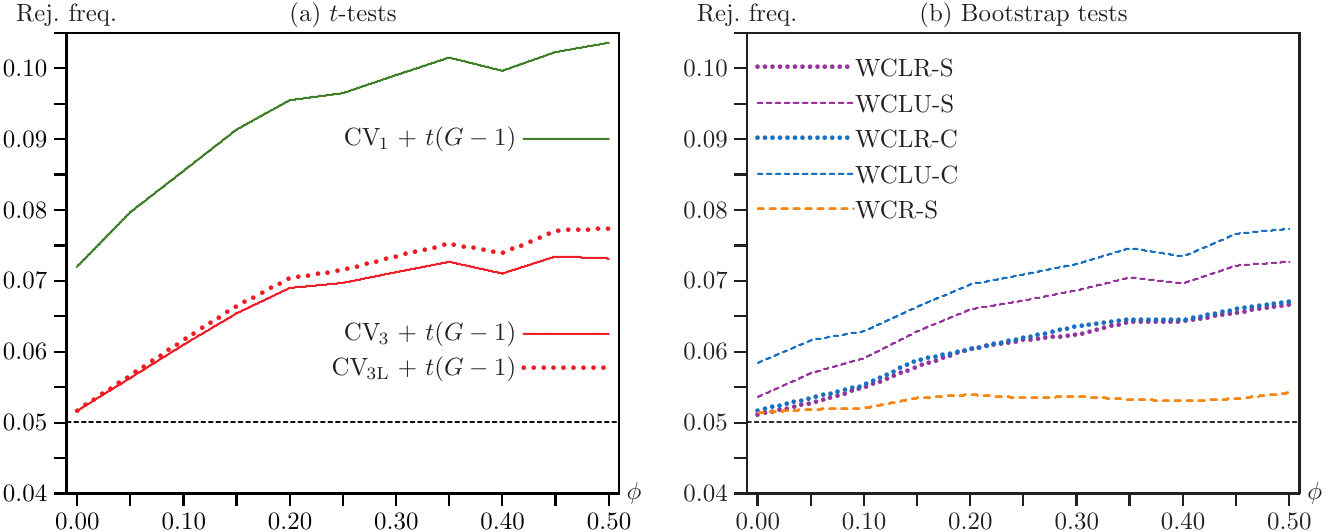}
\end{center}
{\footnotesize
\textbf{Notes:} These experiments are similar to the ones in
\Cref{fig:D}, except that $\phi$ (the parameter that determines how
much intra-cluster correlation there is) varies between 0.0 and~0.5.
As before, $G=24$, $N=12,\tn000$, $\pi=0.31$, $\phi=0.1$, and $k=7$.}
\end{figure}

\Cref{fig:F} deals with the effects of intra-cluster correlation, with
the parameter $\phi$ varying between 0.0 and~0.5 on the horizontal
axes; recall \eqref{eq:ugi} and the discussion around it. In both
panels, all the tests perform worse as $\phi$ increases, which is not
surprising. In Panel~(a), the differences between the two
cluster-jackknife tests become larger as~$\phi$ increases. In
Panel~(b), the WCR-S bootstrap test performs very well, as was also
observed in Panel~(b) of \Cref{fig:D}, and in fact it seems to be
invariant to~$\phi$ for $\phi>0.15$. The other bootstrap tests, on the
other hand, deteriorate noticeably as $\phi$ increases.

\subsection{Variation in number and values of coefficients}

\begin{figure}[tb]
\begin{center}
\caption{Rejection frequencies for tests at the 0.05 level
as functions of $k$}
\label{fig:H}
\includegraphics[width=0.96\textwidth]{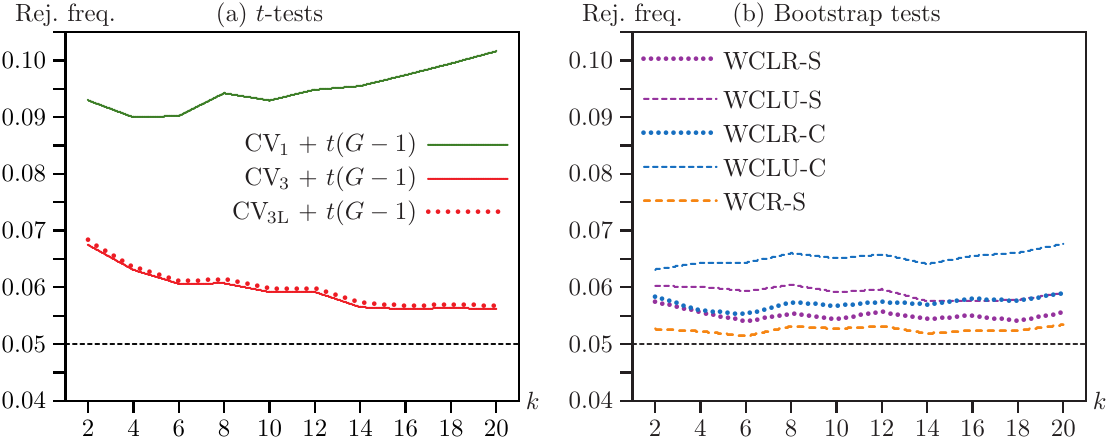}
\end{center}
{\footnotesize
\textbf{Notes:} These experiments are similar to the ones in
\Cref{fig:D}, except that $k$ varies from 2 to 20 by increments of~2.
As before, $G=24$, $N=12,\tn000$, $\pi=0.31$, and $\phi=0.1$.}
\end{figure}

Up to this point, all the models in our experiments have had $k=7$
parameters: a constant term, 6 slope coefficients on regressors of no 
real interest, and the coefficient on the treatment variable. In 
\Cref{fig:H}, we vary $k$ between 2 and~20. In Panel~(a), the 
CV$_{\tn1}$ $t$-test over-rejects and does so a bit more often as $k$ 
increases, while the two cluster-jackknife $t$-tests over-reject to a lesser
degree and do so less often as $k$ increases. In Panel~(b), most of the 
bootstrap tests seem to be almost invariant to~$k$, and they all 
perform very well.

\begin{figure}[tb]
\begin{center}
\caption{Coverage for 95\% confidence intervals as functions of $\beta_k$}
\label{fig:G}
\includegraphics[width=0.96\textwidth]{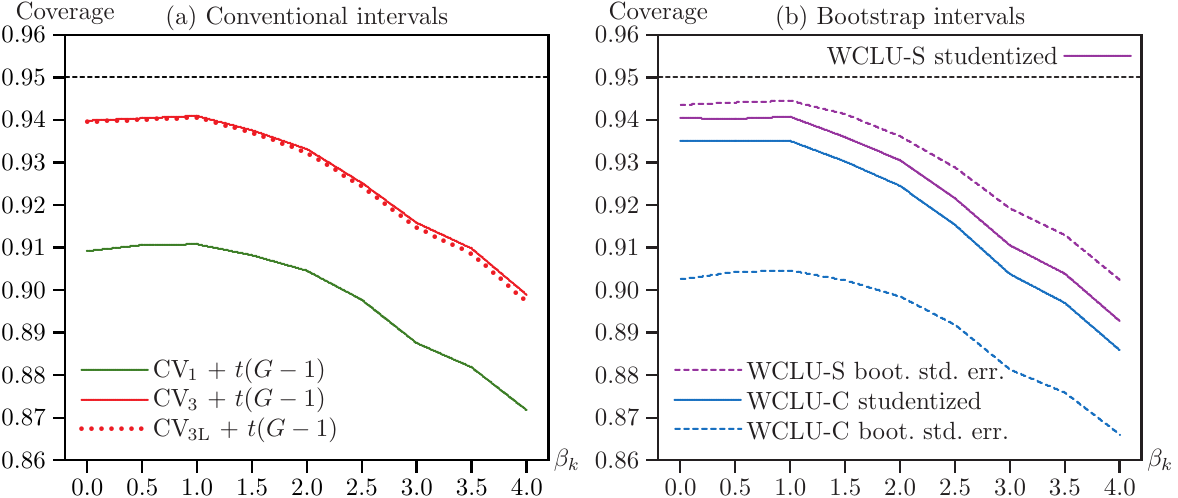}
\end{center}
{\footnotesize
\textbf{Notes:} These experiments are similar to the ones in
\Cref{fig:D}, except that $\beta_k$ varies from 0 to~4, and $\beta_1$
also varies so that $\pi=0.31$. As before, $G=24$, $N=12,\tn000$,
$\phi=0.1$, and $k=7$.}
\end{figure}

In the next set of experiments, we return to confidence intervals. For
hypothesis testing, it is reasonable to impose the null hypothesis
that $\beta_k=0$. But for confidence intervals, $\beta_k$ may often be
non-zero. \Cref{fig:G} shows coverage of 95\% intervals as a function
of $\beta_k$ when $\pi$ is held constant. Note that we do not consider
any variant of the WCU bootstrap here because the LPM does not
estimate the parameter~$\beta_k$. Increasing the value of $\beta_k$
beyond about $\beta_k=1$ steadily reduces coverage. The best interval,
perhaps surprisingly, is the interval \eqref{CIpm} using WCLU-S
standard errors. The conventional intervals based on CV$_{\tn3}$ (or
CV$_{\tn3{\rm L}}$) standard errors perform almost as well. They work 
much better than the usual CV$_{\tn1}$ interval, and there is not much
to choose between them. For large values of~$\beta_k$, coverage is
presumably well below 0.95 because $\E(y_{gi})$ is far from~$0.5$ for
both treated and untreated clusters, although in different directions.
Thus the linear approximation which is at the heart of both standard
asymptotic inference and all of our new methods does not perform
particularly well.

\subsection{Simulation conclusions}

Taken together, our simulation results suggest that the
finite\tkk-sample performance of cluster-robust tests and confidence
intervals for logit models is good to excellent in many cases, but it
can be mediocre when the simulation DGP is extreme in any dimension. 
It seems fairly safe to draw the following conclusions.

\begin{itemize}

\item Conventional $t$-tests based on the CV$_{\tn1}$ variance matrix
and the $t(G-1)$ distribution always over-reject, often severely for
moderate values of~$G$, and the corresponding confidence intervals
always under-cover, often seriously. This method cannot safely be
relied upon. Using the N$(0,1)$ distribution, as popular packages
still do, always makes matters even worse.

\item Cluster jackknife, or CV$_{\tn3}$, $t$-tests always appear to be
more reliable than conventional CV$_{\tn1}$ $t$-tests. However, they
can under-reject moderately in a few cases, and they can over-reject
significantly in others, especially when the fraction of treated
clusters is small, the average value of the dependent variable differs
greatly from one\tkk-half, or the amount of intra-cluster correlation
is large.

\item Linearized cluster jackknife, or CV$_{\tn3{\rm L}}$, standard
errors, which are much cheaper to compute than CV$_{\tn3}$ ones,
usually perform almost as well as the latter. In extreme cases where
they differ, the former tend to be a bit smaller than the latter,
leading to greater over-rejection or under-coverage.

\item The WCLR-S bootstrap often performs well.
When its performance can be distinguished from that of
the WCLR-C bootstrap, it almost always rejects less frequently.

\item All methods can be somewhat unreliable when the binary outcomes are
unbalanced, with most equal to either~0 or~1. This can happen even
when $G$ is quite large.

\item Methods based on the linear probability model, notably the WCR-S
bootstrap, can perform very well indeed. In many cases, the WCR-S and
WCLR-S bootstraps yield similar results.

\item Because confidence intervals based on the WCLR bootstraps are
difficult to compute, we did not study them and cannot recommend them.
Ones based on the WCLU-S bootstrap often perform well, but generally
not much better than conventional cluster-jackknife intervals.
Confidence intervals based on the WCLU-C bootstrap cannot be
recommended.

\item If bootstrap standard errors are desired, they should always be based
on the \mbox{WCLU-S} bootstrap. Surprisingly, it appears that
confidence intervals based on these standard errors may have better
coverage than studentized WCLU-S bootstrap intervals.

\end{itemize}

These conclusions should, of course, be interpreted with caution,
since just how the DGP is specified can substantially affect the
results. For any empirical application, it is always informative to
report the mean of the outcome variable, the number of clusters, the
number of treated clusters (if the regressor of interest is a
treatment dummy), and at least one measure of cluster size variability
\citep*{MNW-influence}. All of those things affect finite\tkk-sample
properties in ways that we have discussed. It may also be desirable to
perform placebo regression experiments, although this may require
quite a bit of effort; see \citet*{BDM_2004},
\citet[Section~3.5]{MNW-guide}, and the next section.

\section{Empirical Examples}
\label{sec:examples}

In this section, we illustrate the tests and confidence intervals that
we have discussed using two empirical examples. The first example has
a relatively small sample ($N=1861$), with a moderate number of
clusters (34) of which about half (16) are treated at the cluster
level. It is thus fairly similar to many of our simulations, and we
expect all the preferred methods to perform well. The second example has
a much larger sample ($N=127,\tn518$), with a small number of clusters
(10), a continuous explanatory variable, and cluster fixed effects.
Because the number of clusters is small, and cluster sizes vary a lot,
we expect different methods to yield substantially different results.

\subsection{Cash Incentives}
\label{sec:cash}

\citet{AL_2009} studies the impact of a randomized cash incentive on
the outcome of a high-stakes examination. A significant sum of money
was offered to ``low-achieving'' students in some Israeli high schools
for passing the exams required to earn their high school matriculation
certificate, or Bagrut. This certificate is a prerequisite for
enrolling in university in Israel. Treatment was assigned randomly at
the school level.

We focus on the estimates for 1861 female students who were enrolled
in $G=34$ schools in the 2001 panel of the study. These are reported
in Table~2, columns 5 and~6, of the original paper. Students were
offered the cash awards in $G_1=16$ of the schools. In addition to the
treatment dummy, the equation includes nine other explanatory
variables, some of which (notably, measures of past performance on
examinations) have considerable explanatory power. Because treatment
was at the school level, school fixed effects cannot be included.

\citet{AL_2009} reports estimates for both the LPM and logit model.
Our results for the former agree with the ones in the paper to the
number of digits reported. Our results for the latter do not quite
agree, however, because the paper reports marginal effects rather than
coefficient estimates. However, the $t$-statistic that is implicitly
reported is within the range of the ones that we obtain.

\citet{AL_2009} reports CV$_{\tn2}$ standard errors for the LPM and
similar ones for the logit model. These are almost certainly more
reliable than CV$_{\tn1}$ standard errors. However, because the number
of clusters is quite small, cluster sizes vary considerably (from 12
to 146), and there is quite a bit of variation in partial leverage
across clusters (see notes to \Cref{tab:cash}), CV$_{\tn3}$ standard
errors are likely to be more reliable than ones based on
CV$_{\tn1}$ or CV$_{\tn2}$ \citep*{MNW-bootknife}.

\begin{table}[tp]
\caption{Effects of Cash Incentives on Passing the Bagrut}
\label{tab:cash}
\vspace*{-0.5em}
\begin{tabular*}{\textwidth}{@{\extracolsep{\fill}}
llccccccc}
\toprule
Model & Method &\textrm{Coef.} &\textrm{Std.\ error}
&\textrm{$t$ stat.} &\textrm{$P$~value} 
&\textrm{CI lower} &\textrm{CI upper} &\textrm{Placebo}\\
\midrule
LPM & CV$_{\tn1}$ &$0.1047$  &$0.0444$ &$2.3572$ &$0.0245$ 
&$\phantom{-}0.0143$ &$0.1952$ &$0.0866$ \\
LPM &  CV$_{\tn2}$ &$0.1047$  &$0.0466$ &$2.2483$ &$0.0314$ 
&$\phantom{-}0.0100$ &$0.1995$ &$0.0681$ \\
LPM &  CV$_{\tn3}$ &$0.1047$ &$0.0506$ &$2.0695$ &$0.0464$ 
&$\phantom{-}0.0018$  &$0.2077$ &$0.0454$ \\
LPM &  WCR-C &$0.1047$ & &$2.3572$ &$0.0393$ &$\phantom{-}0.0055$ &$0.2033$
&$0.0530$ \\
LPM &  WCR-S &$0.1047$ & &$2.3572$ &$0.0418$ &$\phantom{-}0.0042$ &$0.2041$
&$0.0497$ \\
LPM &  WCU-C &$0.1047$ & &$2.3572$ &$0.0381$ &$\phantom{-}0.0064$ &$0.2031$
&$0.0603$ \\
LPM &  WCU-C$^*$ &$0.1047$ &$0.0437$ &$2.3982$ &$0.0223$ &$\phantom{-}0.0159$
&$0.1936$ &$0.0918$ \\
LPM &  WCU-S &$0.1047$ & &$2.3572$ &$0.0401$ &$\phantom{-}0.0053$ &$0.2042$
&$0.0555$ \\
LPM &  WCU-S$^*$ &$0.1047$ &$0.0513$ &$2.0400$ &$0.0494$ &$\phantom{-}0.0003$
&$0.2092$ &$0.0430$ \\
\midrule
Logit &  CV$_{\tn1}$ &$0.7164$ &$0.3149$ &$2.2746$ &$0.0296$
  &$\phantom{-}0.0756$ &$1.3571$ &$0.0794$ \\
Logit &  CV$_{\tn2{\rm L}}$ &$0.7164$ &$0.3303$ &$2.1687$ &$0.0374$
  &$\phantom{-}0.0443$ &$1.3884$ &$0.0607$ \\  
Logit &  CV$_{\tn3}$ &$0.7164$ &$0.3609$ &$1.9850$ &$0.0555$ &$-0.0179$
  &$1.4506$ &$0.0373$ \\
Logit &  CV$_{\tn3{\rm L}}$ &$0.7164$ &$0.3592$ &$1.9941$ &$0.0545$ &$-0.0145$
  &$1.4472$ &$0.0387$ \\
Logit &  WCLR-C &$0.7164$ & &$2.2746$ &$0.0523$ & & &$0.0464$ \\
Logit &  WCLR-S &$0.7164$ & &$2.2746$ &$0.0564$ & & &$0.0426$ \\
Logit &  WCLU-C &$0.7164$ & &$2.2746$ &$0.0457$ &$\phantom{-}0.0151$
&$1.4175$ &$0.0529$ \\
Logit &  WCLU-C$^*$ &$0.7164$ &$0.3095$ &$2.3142$ &$0.0264$
  &$\phantom{-}0.0866$ &$1.3461$ &$0.0846$ \\
Logit &  WCLU-S &$0.7164$ & &$2.2476$ &$0.0487$ &$\phantom{-}0.0042$
&$1.4280$ &$0.0476$ \\
Logit &  WCLU-S$^*$ &$0.7164$ &$0.3645$ &$1.9655$ &$0.0578$ &$-0.0251$
  &$1.4579$ &$0.0364$ \\
\bottomrule
\end{tabular*}
\vskip 6pt {\footnotesize
\textbf{Notes:} There are 1861 observations and 34 clusters. The mean
of the dependent variable is~0.287. The coefficient of variation of
partial leverage across clusters is~0.9655. Two measures of the
effective number of clusters are $G^*(0)=24.3$ and $G^*(1)=14.3$; see
\citet*{CSS_2017} and \citet*{MNW-influence}. Methods based directly on
$t$-statistics use the $t(33)$ distribution. Bootstrap methods use the
Rademacher distribution and $9,\tn999,\tn999$ bootstrap samples so as
to minimize dependence on random numbers. Methods with an asterisk
employ bootstrap standard errors computed using \eqref{bootse} and
$t$-statistics based on them. Methods for which no standard error is
shown use symmetric bootstrap $P$~values based on \eqref{bootps} and
studentized bootstrap confidence intervals based on~\eqref{studboot}.
Entries in the rightmost column are rejection frequencies for placebo
regressions based on $400,\tn000$ replications with $B=999$.}
\end{table}

\Cref{tab:cash} reports several results for a large number of methods.
Of course, we do not recommend reporting this many numbers in
practice. The sixth column shows $P$~values calculated in many
different ways, and the next two columns show the lower and upper
limits of 95\% confidence intervals. For the LPM, all $P$~values are
less than 0.05, and all confidence intervals exclude zero. For the
logit model, every $P$~value is larger than the corresponding one for
the LPM, three of them exceed 0.05, and the three confidence intervals
to which the latter correspond include zero. Overall, there seems to
be modest evidence against the null hypothesis, but the evidence is
much less convincing than we might suppose if we simply looked at the
results for either CV$_{\tn1}$, CV$_{\tn2}$, or CV$_{\tn2{\rm L}}$
standard errors.

The final column of \Cref{tab:cash} contains rejection frequencies for
a placebo regression experiment, where for each replication we add one
additional regressor to the original model and test the hypothesis
that the coefficient on it equals zero. The placebo regressor equals~1
for 16 randomly chosen schools and~0 for the remaining 18 schools.
There are $_{34}\rm{C}_{16} = 2,\tn203,\tn961,\tn430$ ways to choose
the placebo regressor. We did this $400,\tn000$ times and recorded the
fraction of rejections at the 0.05 level.

As can be seen from the last column of \Cref{tab:cash}, several
methods actually under-reject, and no method over-rejects much more
than 9\% of the time. The methods that come very close to 0.05 are the
WCR-S and WCR-C bootstraps for the LPM, and the WCLR-C, WCLU-S, and
WCLU-C bootstraps for the logit model. Interestingly, $t$-tests based
on CV$_{\tn3}$ and CV$_{\tn3{\rm L}}$ both under-reject somewhat. 
Reassuringly, the methods that over-reject most significantly are the 
ones that yield the smallest $P$~values for the actual dataset. These 
$P$~values should evidently not be trusted. Based on all these results, 
we conclude that the true $P$~value for the hypothesis under test is 
probably very close to~0.05.

\subsection{Tuition Fees}
\label{sec:tuition}

There is an extensive literature about the effects of college or
university tuition fees on educational attainment. Many studies have
examined the relationship between tuition and the likelihood of
attending college or attaining a degree; see, for example,
\citet{Heller_1999}.

We examine the effects of tuition fees on scholastic enrollment in
Canada in recent years using data from the public-use version of the
Labour Force Survey (LFS), combined with data on average university
tuition fees in each province. The LFS surveys individuals once per
month, and individuals are included in the survey for six months.
There is much less variation in tuition fees across schools in Canada
than in the United States, because (for the most part) the provinces
regulate them. The tuition data come from Statistics Canada ``Canadian
and international tuition fees by level of study''
Table~37-10-0045-01.

We use the LFS data from 2009\tkk--2019 for males aged 20 and 21 who
reside in one of the ten provinces. The public-use version of the LFS
does not give us the exact age of respondents, so we treat them all as
being the same age. We restrict the sample to the standard Canadian
university academic calendar and therefore omit responses from May
through August. We estimate the following logistic regression at the
individual level:
\begin{equation}
\label{eq:tuition}
{\rm Pr}(\textrm{Student}_{ipt} = 1) = \Lambda\big(\alpha +
\beta\tk\text{Tuition}_{pt} + \text{YEAR}_{\tkk t} + \text{PROV}_{\tn p}
+ \biX_{ipt}\bgamma\big),
\end{equation}
where the outcome variable Student$_{ipt}$ equals~1 if person~$i$ in
province~$p$ in year~$t$ is listed as either a part-time or full-time
student. The regressor of interest is $\text{Tuition}_{pt}$, which is 
the average domestic tuition fee in province~$p$ in year~$t$ expressed 
in thousands of Canadian dollars. Because there are year fixed effects, 
we do not convert these into constant dollars.

The row vector $\biX_{ipt}$ contains two binary variables. One of
these equals~1 when a person lives in any of the nine largest cities
in Canada. We cannot use dummies for different large cities because
each of them is located in only one province. This would make it
impossible to estimate, say, the coefficient on Montreal when a
jackknife sample clustering at the provincial level omits the province
of Quebec. The other dummy variable in $\biX_{ipt}$ indicates whether
someone is a citizen/permanent resident or not. The LFS includes both
permanent residents and citizens, who pay domestic tuition fees, and 
non-permanent residents, who pay international tuition fees. In order 
to minimize the number of individuals who have to pay international 
tuition fees, our sample excludes immigrants who have been in Canada 
for less than ten years. We cluster by province, because our measure 
of tuition fees is constant at the province\tkk-year level and highly 
persistent across years.

We initially estimated the logit model \eqref{eq:tuition} and the
corresponding LPM for men, women, and both together. However, we only
report results for men, because they are the only ones for which the
tuition variable appears to be significant using CV$_{\tn1}$ standard
errors.\footnote{The sample of women contained $120,\tn309$
observations. The tuition coefficient was $-0.0739$ in the logit
model, not much more than half the value of $-0.1302$ for men shown in
\Cref{tab:tuition}. The CV$_{\tn1}$ standard error was slightly larger
(0.0529 instead of 0.0469), and the corresponding $t$-statistic was
therefore much smaller ($-1.3965$ instead of $-2.7745$).} Since our
objective is to illustrate the consequences of using different methods
of inference, we focus on the case where different methods yield
different inferences. There are $127,\tn518$ observations and just ten
clusters. The cluster sample sizes vary from $3,\tn402$ (P.E.I.) to
$37,\tn109$ (Ontario). Thus they vary by a factor of about eleven.
Note that the LFS sample sizes vary much less than actual provincial
populations. For example, as of 2019-Q4, the population of Ontario was
about 93 times the population of P.E.I. The survey includes sampling
weights, which we do not use in this analysis.

\begin{table}[tp]
\caption{Effects of Tuition Fees on Scholastic Enrollment}
\label{tab:tuition}
\vspace*{-0.5em}
\begin{tabular*}{\textwidth}{@{\extracolsep{\fill}}
llccccccc}
\toprule
Model & Method &\textrm{Coef.} &\textrm{Std.\ error}
&\textrm{$t$ stat.} &\textrm{$P$~value} 
&\textrm{CI lower} &\textrm{CI upper} &\textrm{Placebo}\\
\midrule
LPM &  CV$_{\tn1}$ &$-0.0296$  &$0.0106$ &$-2.7899$ &$0.0211$ 
&$-0.0537$ &$-0.0056$ &$0.1332$ \\
LPM &  CV$_{\tn3}$ &$-0.0296$ &$0.0184$ &$-1.6120$ &$0.1414$ 
&$-0.0712$  &$\phantom{-}0.0120$ &$0.0601$ \\
LPM &  WCR-C &$-0.0296$ & &$-2.7899$ &$0.1414$ &$-0.0480$ &$\phantom{-}0.0167$
&$0.0658$ \\
LPM &  WCR-S &$-0.0296$ & &$-2.7899$ &$0.1534$ &$-0.0480$ &$\phantom{-}0.0154$
&$0.0548$ \\
LPM &  WCU-C &$-0.0296$ & &$-2.7899$ &$0.0232$ &$-0.0543$ &$-0.0050$
&$0.1018$ \\
LPM &  WCU-C$^*$ &$-0.0296$ &$0.0101$ &$-2.9405$ &$0.0165$ &$-0.0524$
&$-0.0068$ &$0.1502$ \\
LPM &  WCU-S &$-0.0296$ & &$-2.7899$ &$0.1018$ &$-0.0651$ 
&$\phantom{-}0.0059$ &$0.0747$ \\
LPM &  WCU-S$^*$ &$-0.0296$ &$0.0194$ &$\phantom{-}1.5290$ &$0.1606$
&$-0.0735$ &$\phantom{-}0.0142$ &$0.0508$ \\
\midrule
Logit &  CV$_{\tn1}$ &$-0.1302$ &$0.0469$ &$-2.7745$ &$0.0216$
&$-0.2364$ &$-0.0240$ &$0.1298$ \\
Logit &  CV$_{\tn3}$ &$-0.1302$ &$0.0799$ &$-1.6301$ &$0.1375$
&$-0.3109$ &$\phantom{-}0.0505$ &$0.0574$ \\
Logit &  CV$_{\tn3{\rm L}}$ &$-0.1302$ &$0.0800$ &$-1.6280$ &$0.1380$
  &$-0.3112$ &$\phantom{-}0.0507$ &$0.0575$ \\
Logit &  WCLR-C &$-0.1302$ & &$-2.7745$ &$0.1399$ & & &$0.0639$ \\
Logit &  WCLR-S &$-0.1302$ & &$-2.7745$ &$0.1551$ & & &$0.0527$ \\
Logit &  WCLU-C &$-0.1302$ & &$-2.7745$ &$0.0210$ &$-0.2362$
&$-0.0243$ &$0.0993$ \\
Logit &  WCLU-C$^*$ &$-0.1302$ &$0.0445$ &$-2.9244$ &$0.0169$ &$-0.2310$
  &$-0.0029$ &$0.1464$ \\
Logit &  WCLU-S &$-0.1302$ & &$-2.7745$ &$0.0912$ &$-0.2634$
&$\phantom{-}0.0165$ &$0.0724$ \\
Logit &  WCLU-S$^*$ &$-0.1302$ &$0.0843$ &$-1.5442$ &$0.1569$ &$-0.3210$
  &$\phantom{-}0.0605$ &$0.0485$ \\
\bottomrule
\end{tabular*}
\vskip 6pt {\footnotesize
\textbf{Notes:} There are 127,\tn518 observations and 10 clusters. The
mean of the dependent variable is~0.4208. The coefficient of variation
of partial leverage across clusters is 1.2113, and $G^*(0)=4.575$.
Methods based directly on $t$-statistics use the $t(9)$ distribution.
Bootstrap methods use the six-point distribution of \citet{Webb-6pt}
and $9,\tn999,\tn999$ bootstrap samples so as to minimize dependence
on random numbers. Methods with an asterisk employ bootstrap standard
errors computed using \eqref{bootse} and $t$-statistics based on them.
Methods for which no standard error is shown use symmetric bootstrap
$P$~values based on \eqref{bootps} and studentized bootstrap
confidence intervals based on \eqref{studboot}. Entries in the
rightmost column are rejection frequencies for placebo regressions
based on $400,\tn000$ replications with $B=999$.}
\end{table}

\Cref{tab:tuition} is similar to \Cref{tab:cash}. It reports several
quantities for a large number of methods. One striking feature is how
much $P$~values and confidence intervals vary across methods. Six $P$
values are less than~0.03. These are the ones for the CV$_{\tn1}$
$t$-statistics for both the LPM and logit models, for the WCU-C and
WCLU-C bootstraps, and for $t$-statistics based on bootstrap standard
errors using those two bootstrap methods. At the other extreme, all
the restricted wild bootstrap methods yield $P$~values greater
than~0.135. So do $t$-statistics based on both WCU-S and WCLU-S
bootstrap standard errors.

With only 10 clusters that vary quite a bit in size, and substantial
variation in the partial leverages, it is likely that no method is
very reliable. We attempt to get a sense of which methods work best by
performing a placebo regression experiment, where a placebo regressor
is added to the original model. We generate artificial tuition fee series
by using an AR(1) model, which is simulated separately for each province.
The only parameter that seems to matter is the autoregressive
coefficient. Reported results are for the random walk case, where this
parameter equals~1. For smaller values of this parameter, rejection
frequencies tended to be a little higher.

The rightmost column of \Cref{tab:tuition} shows rejection frequencies
for the coefficient on the placebo regressor based on $400,\tn000$
replications. Because of the fairly large sample size, these
experiments were much more expensive than the comparable experiments
in \Cref{sec:cash}. Computing the CV$_{\tn3}$ variance matrix for the
logit model is by far the most costly part of the process, because it
requires $G$ additional logit estimations. In fact, calculating
CV$_{\tn3}$ takes about 70\% of all the computer time for the placebo
regression experiments of this section. Estimating the LPM and the
original logit model and performing all the bootstrap computations,
with $B=999$, for both models takes only about 30\% of the time.
Remarkably, the cost of calculating CV$_{\tn3{\rm L}}$, which yields
results almost identical to CV$_{\tn3}$ here, is only about 1/41 of
the cost of calculating the latter.

There is evidently a strong, inverse relationship between the placebo
rejection frequencies and the reported $P$~values. That was also the
case for the example of \Cref{sec:cash}. All the methods with
$P$~values less than 0.05 over-reject approximately 10\tkk--15\% of
the time. Conversely, the methods that perform reasonably well all
yield $P$~values greater than~0.13. The methods that perform
particularly well include the WCR-S and WCLR-S bootstraps, along with
\mbox{$t$-tests} based on WCU-S and WCLU-S bootstrap standard errors.
The worst methods for both models are the ones that use $t$-tests
based on either CV$_{\tn1}$ standard errors or WCU-C and WCLU-C
bootstrap standard errors. Interestingly, methods for the logit model
and the LPM that are similar (e.g.\ WCLR-S and WCR-S) tend to perform
almost the same in the placebo regressions.

We conclude that, in sharp contrast to what conventional methods of
inference suggest, there seems to be very limited evidence that average
tuition fees affected scholastic enrollment by men in Canada during
the 2009\tkk--2019 period.

\section{Concluding Remarks}
\label{sec:remarks}

In this paper, we propose several new procedures for inference in
logistic regression models with clustered disturbances. The default
settings in \texttt{R} and \texttt{Stata} use CV$_{\tn1}$ standard
errors combined with critical values from the $\N(0,1)$ distribution,
and our simulations show that the resulting tests can over-reject
severely. Conceptually the simplest of the new procedures is to employ
$t$-tests, or Wald tests, based on the cluster jackknife (CV$_{\tn3}$)
variance matrix, which apparently has not been studied previously in
the context of binary response models, although \texttt{Stata} has
been able to compute it for many years.

We also propose several new procedures based on a linear approximation
to the original nonlinear model, which can be used for a wide variety
of nonlinear models in addition to binary response models. The
simplest procedures involve tests based on the CV$_{\tn3{\rm L}}$
variance matrix, which is just a cluster jackknife matrix for the
linear approximation evaluated at the unrestricted estimates.
Computing CV$_{\tn3{\rm L}}$ can be more than an order of magnitude
cheaper than computing CV$_{\tn3}$ when the number of clusters is not
quite small. In many cases, including both of our empirical examples,
the two variance matrices yield almost identical results. However,
they can yield noticeably different ones when the linear approximation
does not work well.

The other new tests that we propose are variations of the wild cluster
bootstrap. They all start with the same linear approximation as
CV$_{\tn3{\rm L}}$. Conditional on it, they are computationally almost
identical to corresponding variants of the wild cluster bootstrap for
linear regression models. We study four bootstrap tests. Two of these,
denoted WCLR, evaluate the linear approximation at restricted
estimates, and the other two, denoted WCLU, evaluate it at
unrestricted estimates. For each of them, the classic (or ``-C'')
version generates bootstrap samples directly from the cluster-level
empirical scores, and the score (or ``-S'') version generates them
from empirical scores transformed so as to undo some of the
distortions caused by the estimation process, as proposed in
\citet*{MNW-bootknife}.

The WCLR-S/WCLU-S bootstraps employ the usual CV$_{\tn1}$ variance
matrix, not either of the cluster-jackknife ones. It would be much
more expensive to employ the latter, and simulation results for linear
models in \citet*{MNW-bootknife} suggest that, in most cases, doing so
would not lead to better finite\tkk-sample properties.

Extensive simulation experiments, in \Cref{sec:simuls}, suggest that
the new procedures work better, often very much better, than the
conventional approach that uses CV$_{\tn1}$ $t$-tests. However, which
of them works best seems to vary from case to case. CV$_{\tn3}$ and
CV$_{\tn3{\rm L}}$ $t$-tests are always more reliable than CV$_{\tn1}$
$t$-tests. In rare cases, they can even be more reliable than the best
bootstrap tests. The WCLR-S bootstrap often works very well. However,
it can perform poorly when the fraction of 1s in the sample is very
small or very large, and/or when there is a lot of intra-cluster
correlation. In most cases, the WCR-S bootstrap for the linear
probability model rejects less frequently than the WCLR-S bootstrap.
The difference is often tiny, but it can sometimes be substantial,
especially when the latter over-rejects noticeably.

For confidence intervals, WCLU bootstrap methods are much more
convenient than WCLR ones, because there is no need to estimate the
restricted logit model multiple times. The choice between WCLU-C and
WCLU-S is very important, because intervals based on the latter seem
to provide much better coverage with small numbers of clusters. Perhaps
surprisingly, confidence intervals that combine WCLU-S standard errors
with \mbox{$t(G-1)$} critical values often work at least as well as
studentized bootstrap intervals.

Two empirical examples, in \Cref{sec:examples}, demonstrate that our
better methods yield $P$~values and confidence intervals that seem to
be plausible and can differ substantially from conventional ones. The
results of placebo regression experiments are very much in line with
the simulation results of \Cref{sec:simuls}. For both examples, the
methods that over-reject in the placebo regressions always yield lower
$P$~values than the ones that under-reject or reject at about the
correct rate.

\appendix 
\numberwithin{equation}{section} 
\numberwithin{figure}{section} 
\numberwithin{table}{section} 

\makeatletter 
\def\@seccntformat#1{\@ifundefined{#1@cntformat}
  {\csname the#1\endcsname\quad}
    {\csname #1@cntformat\endcsname}}
\newcommand\section@cntformat{}
\makeatother

\section{Appendix A: Probit and Other Binary Response Models}
\label{app:brm}

The methods proposed in this paper can readily be generalized to other
binary response models, based on a symmetric density function
$f(\cdot)$ and corresponding cumulative distribution
function~$F(\cdot)$; things would be a bit more complicated if $f(x)
\neq f(-x)$, but such a density is rarely used. For example, for the
probit model $F(\cdot)$ and $f(\cdot)$ would be the standard normal
cumulative distribution function $\Phi(\cdot)$ and density function
$\phi(\cdot)$, respectively. In the general case, the logistic
function $\Lambda(\cdot)$ in \eqref{eq:lrm} and \eqref{eq:loglik} is
replaced by~$F(\cdot)$.

For general binary response models, \eqref{eq:info} is the inverse of 
the empirical information matrix, which is not necessarily equal to 
minus the inverse of the empirical Hessian. The information matrix 
equality does not hold identically in the sample as it does for the 
logit model. Thus, there are in general two CRVEs for binary response 
models. For both, the filling of the sandwich is the same as in 
\eqref{eq:CV1}, but the scores in \eqref{eq:scores} are replaced by
\begin{equation}
\bis_g(\bbeta) = \sum_{i=1}^{N_g}
\frac{\big(y_{gi} - F(\biX_{gi}\bbeta)\big)
f(\biX_{gi}\bbeta)\biX_{gi}}
{F(\biX_{gi}\bbeta)F(-\biX_{gi}\bbeta)}.
\label{eq:scorebrm}
\end{equation}
The first CRVE is based on the information matrix and is thus given by 
\eqref{eq:CV1}, where the filling of the sandwich is based on 
\eqref{eq:scorebrm} and the bread has \eqref{eq:ups} replaced by
\begin{equation}
\Upsilon_i(\bbeta) = \frac{f^2(\biX_i\bbeta)}
{F(\biX_i\bbeta)F(-\biX_i\bbeta)}\tk.
\label{eq:prups}
\end{equation}
The second CRVE uses the Hessian matrix instead of the information 
matrix as the bread of the sandwich; that is,
\begin{equation}
\mbox{CV$_{\tn1{\rm H}}$:} \qquad \hat\biV_{1{\rm H}}(\hat\bbeta)
= \frac{G}{G-1}\frac{N-1}{N-k}\tk\biH(\hat\bbeta)^{-1}\!
\left(\tk\sum_{g=1}^G \hat\bis_g\hat\bis_g^\top\!\right)\!
\biH(\hat\bbeta)^{-1}.
\label{eq:CV1H}
\end{equation}
For general binary response models, the contribution to the Hessian 
made by the \th{gi} observation depends on the value of~$y_{gi}$. 
Specifically,
\begin{align}
\label{eq:Hesszero}
\biH_{gi}(\bbeta) &= \frac{f'(-\biX_{gi}\bbeta)F(-\biX_{gi}\bbeta)
- f^2(-\biX_{gi}\bbeta)}{F^2(-\biX_{gi}\bbeta)}\biX_{gi}^\top\biX_{gi} 
&\hspace*{-1.5cm}\text{if $y_{gi}=0$},\\
\label{eq:Hessone}
\biH_{gi}(\bbeta) &=
\frac{f'(\biX_{gi}\bbeta)F(\biX_{gi}\bbeta) - f^2(\biX_{gi}\bbeta)}
{F^2(\biX_{gi}\bbeta)}\biX_{gi}^\top\biX_{gi} 
&\hspace*{-1.5cm}\text{if $y_{gi}=1$}.
\end{align}
The $k\times k$ matrices in \eqref{eq:Hesszero} or \eqref{eq:Hessone} 
are summed over all the observations for which $y_{gi}$ equals 0 and~1, 
respectively, to obtain~$\biH(\hat\bbeta)$.

The remainder of \Cref{sec:lrm,sec:linear} are unchanged for general 
binary response models, except that \eqref{eq:infolrm} is replaced by
\begin{equation}
\biJ_g(\bbeta) = \sum_{i=1}^{N_g} \frac{f^2(\biX_{gi}\bbeta)}
{F(\biX_{gi}\bbeta)F(-\biX_{gi}\bbeta)}
\biX_{gi}^\top\biX_{gi}.
\label{eq:infobrm}
\end{equation}

\section{Appendix B: The CV$_{\tn\textbf{2}\mathbf{L}}$ Variance Matrix}
\label{app:CV2L}

The CV$_{\tn2{\rm L}}$ variance matrix can readily be computed by
combining the linearization proposed in \Cref{sec:linear} with the
procedure for calculating CV$_{\tn2}$ given in \citet*{MNW-bootknife},
which is based on an ingenious algorithm proposed in
\citet*{NAAMW_2020}. First, form the $k\times k$ matrices
\begin{equation}
\biA_g = (\hat\biJ^\top\!\hat\biJ)^{-1/2}\hat\biJ_g^\top\!\hat\biJ_g
(\hat\biJ^\top\!\hat\biJ)^{-1/2},
\quad g=1,\ldots,G,
\label{def:AJ}
\end{equation}
where $\biJ_g(\bbeta)$ was defined in \eqref{eq:infolrm}, and
\begin{equation}
\hat\biJ = \sum_{g=1}^G \hat\biJ_g = \sum_{g=1}^G\biJ_g(\hat\bbeta) =
\biX^\top \bUpsilon(\hat\bbeta ) \biX
\label{hatJ}
\end{equation}
is the empirical information matrix. Then calculate the rescaled score
vectors
\begin{equation}
\grave\bis_g = (\hat\biJ^\top\!\hat\biJ)^{1/2}(\bfI_k - \biA_g)^{-1/2}
(\hat\biJ^\top\!\hat\biJ)^{-1/2}\hat\bis_g, \quad g=1,\ldots,G,
\label{eq:bis2f}
\end{equation}
where $\hat\bis_g = \bis_g(\hat\bbeta)$, and $\bis_g(\bbeta)$ was
defined in \eqref{eq:scores}. The variance matrix we want is then
\begin{equation}
\mbox{CV$_{\tn2{\rm L}}$:}\qquad \hat\biV_{2{\rm L}}(\hat\bbeta) =
(\hat\biJ^\top\!\hat\biJ)^{-1}\Big(\tk\sum_{g=1}^G \grave\bis_g
\grave\bis_g^\top\Big)(\hat\biJ^\top\!\hat\biJ)^{-1}.
\end{equation}
CV$_{\tn2{\rm L}}$ looks very similar to CV$_{\tn1}$ given in
\eqref{eq:CV1}. It just omits the leading scalar factor and replaces
the $\hat\bis_g$ by the $\grave\bis_g$ given in \eqref{eq:bis2f}.

\section{Appendix C: The \texttt{logitjack} Package}
\label{app:ljack}

\def\hangpara{\hangindent=\parindent\hangafter=1\noindent}

We have developed a \texttt{Stata} package called \texttt{logitjack}
that computes the CV$_{\tn3{\rm L}}$ and (optionally) CV$_{\tn3}$
variance matrices and performs the WCLR-C, WCLR-S, \mbox{WCLU-C}, and
\mbox{WCLU-S} bootstraps. The latest version may be obtained from
\url{https://github.com/mattdwebb/logitjack}. Alternatively, logitjack
is available on \texttt{Stata}'s \texttt{SSC} server. The data and 
programs used in the paper may be found at
\url{http://qed.econ.queensu.ca/pub/faculty/mackinnon/logitjack/}.

\subsection{Syntax}

The syntax for \texttt{logitjack} is

\smallskip

\texttt{
	\hangpara logitjack varlist, \underbar{clus}ter(varname)
	[\underbar{fevar}(varlist) \underbar{boot}strap \underbar{no}null}\par 
\texttt{
	\hangpara \texttt{reps}(\#) \texttt{\underbar{jack}knife} \underbar{sam}ple(string)]
}

\smallskip

\noindent Here \texttt{varlist} contains a list of variables. The
first one is the dependent variable, the second is the regressor for
which standard errors and $P$~values are to be calculated, and the
remaining ones are all the other continuous and binary regressors.
Categorical variables to be treated as fixed effects should be listed 
using the \texttt{\underbar{fevar}} option.

\smallskip

\texttt{\underbar{clus}ter(varname)} is mandatory, where
\texttt{varname} is the name of the variable by which the observations
are clustered. For every observation, it should equal one of $G$
positive integers.

\smallskip

\texttt{\underbar{fevar}(varlist)}. Categorical variables to be
included in the model as fixed effects should be listed here. They are
handled equivalently to \texttt{i.varlist} in a logit model. Since
this option uses a generalized inverse, CV$_{\tn3}$ can be calculated 
even when some of the omit-one-cluster subsamples are singular.
This always happens with cluster-level fixed effects. In contrast, the
\texttt{Stata} command \texttt{jackknife:\ logit y x i.clustervar,
cluster(clustervar)} is unable to estimate CV$_{\tn3}$. It drops every
subsample because each contains a different fixed effect which
is not estimable.

\smallskip

\texttt{\underbar{boot}strap} requests that bootstrap $P$~values be
computed. The default number of bootstraps is~999. This can be
changed using the \texttt{reps}(\#) option. The weight distribution
used depends on the number of clusters. When there are 13 or more
clusters, Rademacher weights are used. When there are 12 or fewer
clusters, \citet{Webb-6pt} weights are used. This option requests
restricted versions of the wild cluster bootstrap. The \texttt{nonull}
option instead requests unrestricted versions.

\smallskip

\texttt{\underbar{no}null} specifies that the bootstrap DGP should be
unrestricted. When it is specified, the package displays bootstrap
standard errors, confidence intervals, and $P$~values, based on both
the WCLU-C and WCLU-S bootstraps. This option has the same effect
whether it is used alone or in addition to the
\texttt{\underbar{boot}strap} option.

\smallskip

\texttt{reps}(\#) allows the number of bootstrap replications to be
specified. When it is not invoked, the
\texttt{\underbar{boot}strap} and \texttt{\underbar{no}null} options
both default to 999 replications. If this option is invoked in 
isolation, then restricted versions of the bootstrap are calculated, as
if \texttt{boot} had been specified without \texttt{\underbar{no}null}.

\smallskip

\texttt{\underbar{jack}knife} requests calculation of the CV$_{\tn3}$ 
standard error. This is an option because CV$_{\tn3}$ is relatively 
expensive. The CV$_{\tn1}$ and CV$_{\tn3{\rm L}}$ standard errors are 
always calculated. This option is useful when CV$_{\tn3}$ is desired 
but the inclusion of cluster-level fixed effects causes issues for
\texttt{Stata}'s \texttt{jackknife} prefix. 

\smallskip

\texttt{\underbar{sam}ple(string)} limits the sample. Use the text you would
enter after an ``if'' in a regression command. For instance,
\texttt{sample(female==1)} is equivalent to ``\texttt{if female==1}.''

\subsection{Illustration}

In the remainder of this appendix, we illustrate the use of
\texttt{logitjack} with an example that employs the \texttt{webuse}
dataset \texttt{nlswork}. The objective is to predict whether a person
is a college graduate. The variable of interest is a dummy variable
indicating that the person is from a southern state. There is
clustering by industry, with just twelve industries.

The first commands load and clean the dataset.

\begin{verbatim}	
    webuse nlswork, clear
    gen age2 = age*age
    drop if race==3
    drop if inlist(ind,41,54)
    gen white = race==1
\end{verbatim}

\noindent For comparison purposes, the native Stata logit estimate is
obtained from the command

\medskip

\noindent\texttt{logit collgrad south msp white union ln\_wage age age2
i.ind, cluster(ind)}

\medskip

\noindent It yields the results

\begin{verbatim}
Logistic regression                                     Number of obs = 18,919
Wald chi2(7)  =      .
Prob > chi2   =      .
Log pseudolikelihood = -6873.2595                       Pseudo R2     = 0.2622
\end{verbatim}

\begin{verbatim}
(Std. err. adjusted for 12 clusters in ind_code)
------------------------------------------------------------------------------
             |               Robust
    collgrad | Coefficient  std. err.      z    P>|z|     [95% conf. interval]
-------------+----------------------------------------------------------------
       south |   .3468109   .1905475     1.82   0.069    -.0266554    .7202773
\end{verbatim}

\medskip

\noindent The simplest \texttt{logitjack} command for this model is

\medskip

\noindent \texttt{logitjack collgrad south msp white union ln\_wage,
cluster(ind) fevar(ind)}\par

\medskip

\noindent The resulting output is:

\begin{verbatim}
Jackknife cluster statistics for binary response models.
Estimates for south when clustered by ind_code.
There are 18919 observations within 12 ind_code clusters.
Logistic Regression Output
		
  s.e. |      Coeff   Sd. Err.   t-stat  P value    CI-lower    CI-upper
-------+----------------------------------------------------------------
   CV1 |   0.346811   0.190638   1.8192   0.0962   -0.072781    0.766403
  CV3L |   0.346811   0.303466   1.1428   0.2774   -0.321113    1.014735
------------------------------------------------------------------------
	\end{verbatim}
	
\begin{verbatim}
Cluster Variability

 Statistic |       Ng    Lin beta no g
-----------+--------------------------
       min |    38.00         0.050280
        q1 |   153.50         0.333767
    median |   987.00         0.356937
      mean |  1576.58         0.336269
        q3 |  2318.00         0.376996
       max |  6247.00         0.433176
-----------+--------------------------
   coefvar |     1.19         0.282305
\end{verbatim}

\noindent Adding the \texttt{\underbar{jack}knife} option adds an
additional row to the first table and an additional column to the second.

\medskip

\noindent \texttt{logitjack collgrad south msp white union ln\_wage,
cluster(ind) fevar(ind) jack}\par

\begin{verbatim}	
Logistic Regression Output
		
  s.e. |      Coeff   Sd. Err.   t-stat  P value    CI-lower    CI-upper
-------+----------------------------------------------------------------
   CV1 |   0.346811   0.190638   1.8192   0.0962   -0.072781    0.766403
   CV3 |   0.346811   0.295580   1.1733   0.2654   -0.303757    0.997379
  CV3L |   0.346811   0.303466   1.1428   0.2774   -0.321113    1.014735
------------------------------------------------------------------------
\end{verbatim}

\begin{verbatim}
Cluster Variability
	
 Statistic |       Ng      Lin beta no g   beta no g
-----------+----------------------------------------
       min |    38.00           0.050280    0.059133
        q1 |   153.50           0.333767    0.333777
    median |   987.00           0.356937    0.356958
      mean |  1576.58           0.336269    0.337106
        q3 |  2318.00           0.376996    0.377489
       max |  6247.00           0.433176    0.432746
-----------+----------------------------------------
   coefvar |     1.19           0.282305    0.274484
\end{verbatim}

\noindent The next command calculates restricted wild bootstrap
$P$~values with the default number of replications.

\medskip

\noindent \texttt{logitjack collgrad south msp white union ln\_wage,
cluster(ind) fevar(ind) boot}\par

\begin{verbatim}
Restricted Bootstrapped Linearized Regression Output

      WCLR |      Coeff   Sd. Err.   t-stat  P value
-----------+----------------------------------------
   CLASSIC |   0.346811   0.190638   1.8192   0.4565
     SCORE |   0.346811   0.190638   1.8192   0.3774
----------------------------------------------------
P-values calculated with 999 replications and Webb weights.
\end{verbatim}

\noindent The following command is essentially the same as the last
one, but it specifies an alternate number of replications.

\medskip

\noindent \texttt{logitjack collgrad south msp white union ln\_wage,
cluster(ind)///}\par
\noindent \texttt{fevar(ind) reps(1999)}\par

\begin{verbatim}
Restricted Bootstrapped Linearized Regression Output

      WCLR |      Coeff   Sd. Err.   t-stat  P value
-----------+----------------------------------------
   CLASSIC |   0.346811   0.190638   1.8192   0.4777
     SCORE |   0.346811   0.190638   1.8192   0.4122
----------------------------------------------------
P-values calculated with 1999 replications and Webb weights.
\end{verbatim}

\noindent The next command estimates unrestricted wild bootstrap
$P$~values and confidence intervals with the default number of
replications.

\medskip

\noindent{\texttt{logitjack collgrad south msp white union ln\_wage,
cluster(ind) ///}\par
\noindent{\texttt{fevar(ind) nonull}\par

\begin{verbatim}
	Unrestricted Bootstrapped Linearized Regression Output

      WCLU |      Coeff   Sd. Err.   t-stat  P value
-----------+----------------------------------------
   CLASSIC |   0.346811   0.190638   1.8192   0.3323
     SCORE |   0.346811   0.190638   1.8192   0.3854
----------------------------------------------------
P-values calculated with 999 replications and Webb weights.
\end{verbatim}

\begin{verbatim}
Unrestricted Bootstrapped Confidence Intervals

          WCLU |      Coeff    std.er.      WCLU CI-low       WCLU CI-up
---------------+--------------------------------------------------------
CLASSIC-CV1-se |   0.346811   0.190638          -0.4316           1.1428
 CLASSIC-WB-se |   0.346811   0.183550          -0.0572           0.7508
---------------+--------------------------------------------------------
  SCORE-CV1-se |   0.346811   0.190638          -0.5141           1.2153
   SCORE-WB-se |   0.346811   0.316932          -0.3508           1.0444
------------------------------------------------------------------------
\end{verbatim}

In this example, the default $P$~value from native \texttt{Stata}, using
the $\N(0,1)$ distribution, is~0.069. Because $G$ is only 12 and cluster
sizes vary greatly, this is much too small. Using any of the procedures 
described in this paper changes inferences noticeably. For instance, the
CV$_{\tn3{\rm L}}$ and CV$_{\tn3}$ $P$~values are both over 0.25, and the
bootstrap $P$~values are all above~0.30.

\bibliography{mnw-logit}
\addcontentsline{toc}{section}{\refname}

\end{document}